\PassOptionsToPackage{numbers,sort&compress}{natbib}  
\documentclass[10pt,amsmath,twocolumn,notitlepage,preprintnumbers,amssymb,nofootinbib,superscriptaddress]{elsarticle}

\usepackage[margin=1in]{geometry}% less space at margins

\journal{Advances in Space Research}

\makeatletter
\def\ps@pprintTitle{%
 \let\@oddhead\@empty
 \let\@evenhead\@empty
 \def\@oddfoot{\footnotesize\itshape
  Preprint submitted to \@journal \hfill
  FERMILAB-PUB-25-0354-PPD} %\today}%
 \let\@evenfoot\@oddfoot}
\makeatother

\usepackage{times}
\usepackage{setspace}

\usepackage{graphicx}
\usepackage{float}
\usepackage{subcaption}
\usepackage{booktabs}
\usepackage{tabularx}
\usepackage{enumitem}

\usepackage{amsmath}
\usepackage{amssymb}

\usepackage[compact]{titlesec}  

\renewcommand{\thesection}{\arabic{section}}
\renewcommand{\thesubsection}{\thesection.\arabic{subsection}}
\renewcommand{\thesubsubsection}{\thesubsection.\arabic{subsubsection}}

\titleformat{\section}
  {\normalfont\large\bfseries}{\thesection}{1em}{} % Sections: Large, bold, with proper numbering

\titleformat{\subsection}
  {\normalfont\normalsize\bfseries}{\thesubsection}{1em}{} % Subsections: Normal size, bold

\titleformat{\subsubsection}
  {\normalfont\normalsize\itshape}{\thesubsubsection}{1em}{} % Subsubsections: Normal size, italic

\titlespacing{\section}{0pt}{10pt}{5pt}
\titlespacing{\subsection}{0pt}{8pt}{4pt}
\titlespacing{\subsubsection}{0pt}{6pt}{3pt}

\usepackage{caption}
\usepackage{url}
\usepackage[colorlinks=true]{hyperref}

\usepackage{xspace}

\usepackage{color}
\definecolor{violet}{rgb}{0.5,0,1}
\definecolor{orange}{rgb}{1,0.5,0}
\definecolor{teal}{rgb}{0,0.5,0.5}
\definecolor{burntorange}{rgb}{0.8, 0.33, 0.0}
\newcommand{\unit}[1]{\ensuremath{\mathrm{\,#1}}\xspace}

\newcommand{\e}{\unit{e^{-}}}

\begin{document}
%\linenumbers

\begin{frontmatter}

\onecolumn{
\title{\boldmath DarkNESS: A skipper-CCD NanoSatellite for Dark Matter Searches}

%First author
\author[UIUC]{Phoenix Alpine\corref{cor1}}
%\corref{cor1}}
\ead{alpine2@illinois.edu}
\cortext[cor1]{Corresponding Author}

%More authors, linking to affiliations below.
\author[UIUC]{Samriddhi Bhatia}

\author[FNAL]{Ana M. Botti}

\author[FNAL]{Brenda A. Cervantes-Vergara}

\author[FNAL]{Claudio R. Chavez}

\author[IIIE]{Fernando Chierchie}

\author[FNAL,UC]{Alex Drlica-Wagner}

\author[SBU]{Rouven Essig}

\author[FNAL,UC]{Juan Estrada}

\author[TAU]{Erez Etzion}

\author[FNAL]{Roni Harnik}

\author[UC]{Terry Kim}

\author[UIUC]{Michael Lembeck}

\author[UIUC]{Qi Lim}

\author[GSFC]{Bernard J. Rauscher}

\author[FNAL,KICP]{Nathan Saffold\corref{cor1}}
\ead{nsaffold@fnal.gov}

\author[FNAL]{Javier Tiffenberg}

\author[FNAL]{Sho Uemura}

\author[SBU]{Hailin Xu}

%Addresses for affiliations
\address[UIUC]{University of Illinois, Urbana-Champaign, IL, 61801, USA}
\address[FNAL]{Fermi National Accelerator Laboratory, Batavia, IL 60510, USA}
\address[IIIE]{Instituto de Inv. en Ing. El\'ectrica ``Alfredo C. Desages'' (IIIE) CONICET, Bah\'ia Blanca, Argentina}
\address[UC]{Department of Astronomy and Astrophysics, University of Chicago, Chicago, IL 60637, USA}
\address[SBU]{C.N.~Yang Institute for Theoretical Physics, Stony Brook University, Stony Brook, NY 11794, USA}
\address[TAU]{School of Physics and Astronomy, Tel Aviv University, Tel Aviv, 69978, Israel}
\address[KICP]{Kavli Institute for Cosmological Physics, University of Chicago, Chicago, IL 60637, USA}
\address[GSFC]{NASA Goddard Space Flight Center, Greenbelt, MD, USA}

% \address[ATA]{ATA Engineering, Berkeley, CA 94704, USA}

%\date{\today}
%\preprint{FERMILAB-FN-1265-PPD}

%\maketitle
\begin{abstract}

The Dark matter Nanosatellite Equipped with Skipper Sensors (DarkNESS) deploys a recently developed skipper-CCD architecture with sub-electron readout noise in low Earth orbit (LEO) to investigate potential signatures of dark matter (DM). The mission addresses two interaction channels: electron recoils from strongly interacting sub-GeV DM and X-rays produced through decaying DM. Orbital observations avoid attenuation that limits ground-based measurements, extending sensitivity reach for both channels. The mission proceeds toward launch following laboratory validation of the instrument. A launch opportunity has been secured through Firefly Aerospace’s DREAM 2.0 program, awarded to the University of Illinois Urbana-Champaign (UIUC). This will constitute the first use of skipper-CCDs in space and evaluate their suitability for low-noise X-ray and single-photon detection in future space observatories.

\end{abstract}
}

\end{frontmatter}
%\maketitle

\twocolumn
\section{Mission Motivation}
\label{sec:mission}

Understanding the nature of dark matter (DM), a non-luminous component of the cosmic mass-energy content, remains a central challenge in contemporary physics~\citep{Planck_2020,Arbey_2021}. While direct detection efforts have long focused on Weakly Interacting Massive Particles (WIMPs)~\citep{Akerib_2022_snowmass}, repeated null results have motivated the search for a broader class of DM candidates. Most DM searches operate underground to suppress cosmogenic backgrounds, but the Earth's atmosphere and crust attenuate X-rays and certain DM signatures, particularly DM that interacts with a large cross section.

Charge-Coupled Devices (CCDs) have a demonstrated heritage in space-based imaging~\citep{GalileoCCD_1979,Trauger_HST_1990}. Fermi National Accelerator Laboratory's (FNAL) `skipper' amplifier is a recent advancement in CCD technology, that capacitively couples a floating-gate amplifier to the CCD's sense node,  allowing repetitive non-destructive readout of the pixel charge~\cite{Tiffenberg:skipper2017}.  With a sufficient number of skipper samples, sub-electron noise can be achieved, significantly improving sensitivity to rare, low energy ionization events. Skipper-CCDs have already demonstrated laboratory sensitivity to X-rays from hypothetical DM decay and electron recoils from sub-GeV DM scattering~\citep{Sensei2023,SENSEI1e_2025}.

% Science-grade CCDs with conventional amplifiers are limited a noise performance of about 2\,\e of readout noise. skipper-CCDs are able to achieve sub-electron noise by employing RNDR on the pixel charge measurement, offering sub-electron noise performance. This reduced noise performance allows low-threshold rare event searches, with skipper-CCDs providing world-leading sensitivity to sub-GeV DM that interacts with electrons~\cite{Sensei2023,SENSEI1e_2025,DAMIC-M_2025}. Along with their sensitivity to sub-GeV DM interactions, skipper-CCDs have demonstrated Fano-limited energy resolution in the X-ray regime, providing a useful tool to perform X-ray spectroscopy~\cite{Sensei2023}.

The Dark matter Nanosatellite Equipped with Skipper Sensors (DarkNESS) is the first orbital mission designed to operate skipper-CCDs in the space environment. As a scientific observatory, it aims to probe DM signatures that are inaccessible from ground-based instruments. The mission is configured to perform targeted DM searches while also serving as a technology demonstration for future low-noise space instrumentation developed by FNAL. Operating above the atmosphere avoids the shielding effects that limit Earth-based experiments, permitting observation of X-rays and strongly interacting DM models that are attenuated in the Earth's atmosphere before reaching ground-based instruments.

% Order of Objectives, same words, re-organized to match as presented in Section 2:
DarkNESS pursues two complementary science goals. The first focuses on identifying strongly interacting sub-GeV DM via low-energy ionization signals, exploiting the skipper-CCD's sub-electron noise characteristics to measure the low-energy ionization rate throughout the orbit environment. The second is to search for faint X-rays in the 1–10\,keV band that could originate from decaying DM in the Galactic DM halo. These measurements probe previously inaccessible DM parameter space.

% DarkNESS pursues two complementary science goals. The first is to search for faint X-rays in the 1–10\,keV band that could originate from decaying DM in the Galactic DM halo. The second focuses on identifying strongly interacting sub-GeV DM via low-energy ionization signals, exploiting the skipper-CCD's sub-electron noise characteristics to measure the low-energy ionization rate throughout the orbit environment. These measurements probe previously inaccessible DM parameter space.

The scientific instrument—integrated by FNAL and tested in collaboration with the University of Illinois at Urbana-Champaign (UIUC)—completed environmental testing to evaluate performance under simulated thermal conditions. The mission passed Critical Design Review (CDR) and is proceeding toward launch under the Firefly Aerospace DREAM 2.0 program, awarded to UIUC. This paper presents the DarkNESS science objectives, system configuration, mission feasibility, and the trade space for deploying low-threshold skipper-CCDs within the mass, volume, and power constraints imposed by NanoSatellite-class observatories.

\section{Scientific Objectives}
\label{sec:science}
Atmospheric attenuation conceals certain signatures of DM, requiring space-based instrumentation to probe these models. The DarkNESS mission will search for two such DM signatures from LEO, with sensitivity to strongly interacting sub-GeV DM and X-rays from decaying DM. In addition to the scientific objectives, DarkNESS will qualify the skipper-CCD technology in the space environment, thereby advancing the Technology Readiness Level (TRL) of space-based skipper-CCDs.

\subsection{Strongly interacting sub-GeV Dark Matter}
% Spelling and Grammar Checked

DM direct detection experiments are typically performed in underground laboratories to shield the detectors from cosmogenic radiation and reach sensitivities required for rare-event searches involving extremely small interaction cross-sections. However, if DM interacts strongly with ordinary matter, it will scatter in the Earth’s atmosphere and crust, attenuating the DM flux reaching terrestrial detectors. At large interaction cross sections, the flux would not reach detectors deployed underground or on the Earth's surface. This scenario was examined in detail for sub-GeV DM in~\citep{Emken:2019tni} and showed that strongly interacting sub-GeV DM coupling to the Standard Model through an ultralight dark photon mediator could be a subdominant component ($f_{\chi} \lesssim 0.1\%$) of the cosmological DM. 

\begin{figure}[t]
    \centering
    \includegraphics[width=\columnwidth]{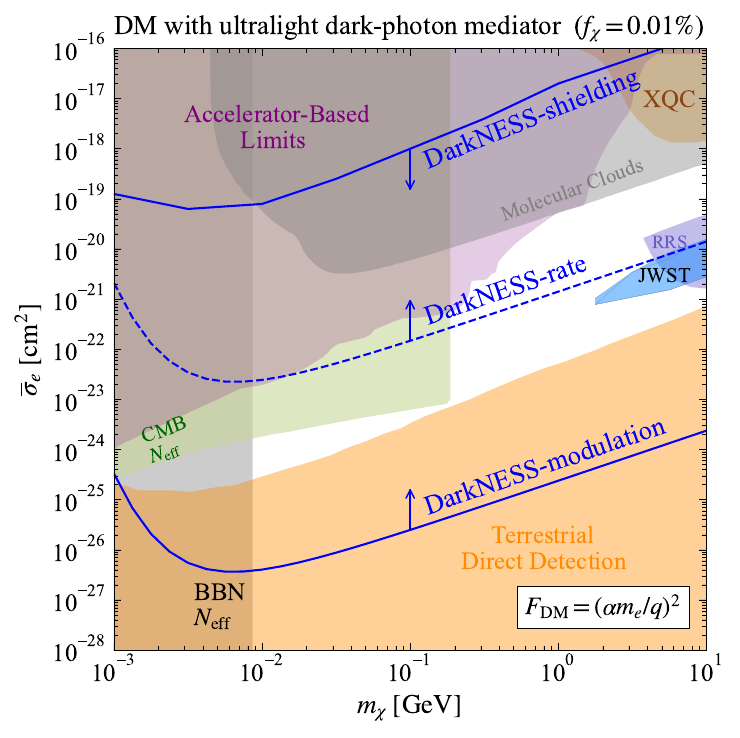}
\caption{Dark-matter electron interaction parameter space showing recent bounds on the DM-electron scattering cross section for a subdominant component of DM ($f_\chi=0.01\%$). DarkNESS will help expand the current direct detection exclusion bands (orange contour) up to higher cross-sections, constraining the parameter space for strongly interacting sub-GeV DM. The DarkNESS discovery reach is between the solid blue line of `DarkNESS-shielding' and `DarkNESS-modulation'. The dashed blue curve labeled `DarkNESS-rate' is where the DM-electron interaction rate equals the assumed background rate, while much lower cross sections can be probed if one searches for a modulation signal. This plot assumes that DM interacts via an ultralight mediator, an exposure of 0.1~gram-months, and a background of $10^9$ events. Other limits shown are from~\cite{Du:2024afd, Creque-Sarbinowski:2019mcm, Munoz:2018pzp,Prinz:1998ua,Ball:2020dnx,Magill:2018tbb,ArgoNeuT:2019ckq,Plestid:2020kdm,Mahdawi:2018euy,Rich:1987st,Prabhu_2023}.}
\label{fig:Sat-sensitivity-1}
\label{fig:DarkNESS-Science}
\end{figure}

DarkNESS will use skipper-CCDs with sub-electron noise to search for strongly interacting sub-GeV DM. Observations target the constellation Cygnus, which lies near the Solar Apex—the direction of the Sun’s motion through the Galaxy’s hypothesized DM halo. This relative motion creates a \textit{dark matter wind}, with particles appearing to flow from Cygnus' direction~\citep{Drukier_1986,Freese_2013}. Aligning the instrument toward the wind increases an expected DM flux. As the NanoSatellite orbits Earth, planetary shadowing is expected to result in a modulated signature in the instrument’s low-energy event rate~\citep{Kouvaris_2014}.

%\re{
The DarkNESS instrumentation package includes an active mass of approximately 2~grams. Considering an orbital modulation search using approximately 450 observations of Cygnus with a 10-minute exposure time and selecting 50\% of the exposed pixels after masking high-energy hits, DarkNESS can obtain a 0.1~gram-month exposure to achieve the discovery reach (5$\sigma$) shown with the lower solid blue line of Fig.~\ref{fig:Sat-sensitivity-1}.  
This assumes a fractional DM abundance of $f_\chi = 0.01\%$ and $10^9$ background events (constant in time). Along the dashed blue line, we show the discovery reach from requiring that the total number of DM events is larger than the number of background events, showing that a modulation search can perform much better.  The DarkNESS sensitivity will expand the current upper limits on the DM-electron scattering cross section to higher cross-section values, filling the gap above terrestrial direct-detection searches, represented by the orange shaded region. The reach of the DarkNESS upper limit is constrained by the amount of shielding between the skipper-CCDs and the incoming DM flux, and the DarkNESS design uses minimal shielding ($\sim$50\,nm Al) to block stray light while maintaining sensitivity to strongly interacting sub-GeV DM. In order to produce a measurable signal, the DM has to pass through the shield ($\sim$50\,nm Al) and CCD gate structures ($\sim$1\,$\mu$m Si) into the active region of the skipper-CCD. The upper blue line in Fig.~\ref{fig:Sat-sensitivity-1} conservatively assumes that the DM has to pass through a total shielding of $10~\mu$m of silicon. 
%} 

\subsection{X-ray Signatures of Dark Matter}

% -Sep23 2024
%Nate's version:
DM decay is generically predicted in various particle DM scenarios, and photons are one of the promising channels to probe DM decay~\citep{Boddy_2022}. DM decay processes could give rise to a monoenergetic photon line, providing an indirect detection signature to search for DM. Space-based observations are necessary to probe $\mathcal{O}\mathrm{(keV-GeV)}$ photons, since at these energies they do not penetrate the Earth's atmosphere.

\begin{figure}[t]
    \centering
    \includegraphics[width=\columnwidth]{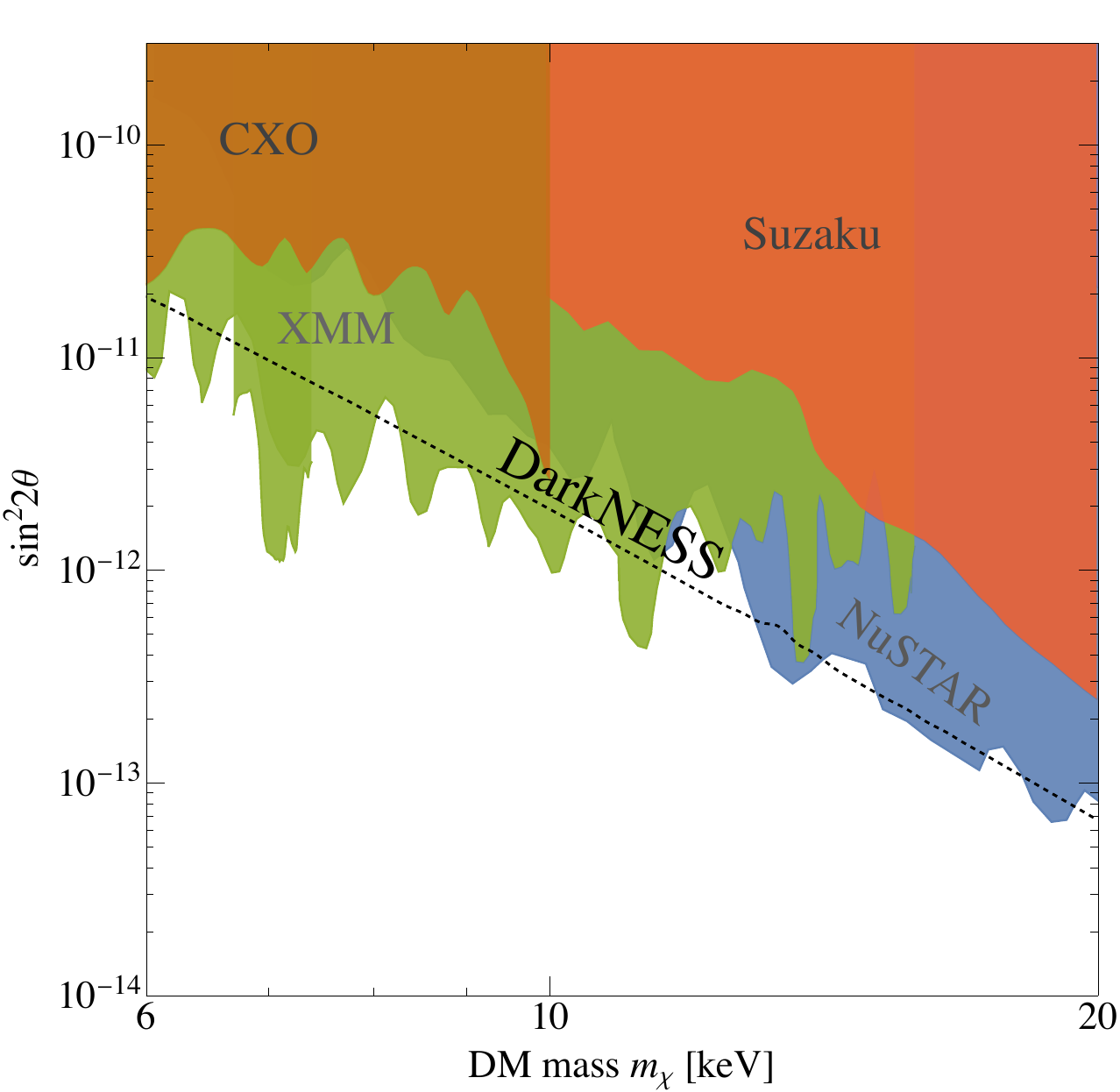}
\caption{Exclusion limits on the DM decay rate into X-rays cast in units of $\sin^2 2\theta$ where $\theta$ is the mixing angle between the sterile and active neutrino. Black dashed line shows the projected DarkNESS 90\% C.L. upper limit using 25 hours (90\,ks) of exposure time, assuming a Galactic Center background model from Ref.~\cite{figueroa}. The DarkNESS limit is compared with the best current limits from XMM in green~\citep{XMM,Dessert_2020}, NuSTAR in blue~\citep{NuStar}, Suzaku in red-orange~\citep{Tamura_2015}, and CXO in orange~\citep{Sicilian_2020}.\label{fig:Sat-sensitivity-Xray}}
\label{fig:DarkNESS-ScienceXray}
\end{figure}

To probe decaying DM, the DarkNESS mission will search for unidentified X-ray lines in observations of the Galactic Center. During its operational lifetime, DarkNESS will have the opportunity to perform over 600 observations of the Galactic Center, each with a 15-minute integration time that can be split into shorter exposures to mitigate background radiation. This will result in a total exposure time of approximately 500\,ks. The long exposure and wide Field-of-View (FOV) provide a large predicted signal rate from benchmark decaying DM models. For example, a sterile neutrino with $\mathcal{O}$(keV) mass decaying into an X-ray photon and an active neutrino arises from well-motivated extensions of the Standard Model and could make up the cosmological DM~\citep{Adhikari_2017}.

The decay of a ${\sim}$7\,keV sterile neutrino could explain an unidentified X-ray line at 3.5\,keV that was detected with high significance using stacked observations of galaxy clusters from instruments on the XMM-Newton satellite~\citep{Bulbul_2014}. This detection inspired a flurry of follow-up observations with mixed results (e.g.~\cite{Boyarsky:2014jta, Abazajian:2001vt, Abazajian:2014gza, Boyarsky:2018ktr}).
There is still some debate on how to interpret these conflicting results. The most recent results in this area disfavor the interpretation of the 3.5~keV line as resulting from DM decay (e.g.~\cite{Dessert_2020, Dessert:2023fen, Boyarsky_2020,2022swift, 2024TanDekkerDrlica}) and suggest that the line was always an artifact rather than a valid signal. DarkNESS will help resolve this discrepancy by making dedicated observations of the Galactic Center's diffuse X-ray background, providing a large new dataset to search for unidentified spectral lines due to DM decays. Using a 90\,ks exposure, the expected DarkNESS sensitivity to sterile neutrino DM decaying to X-rays is shown in Fig.~\ref{fig:Sat-sensitivity-Xray}.

%The optimal exposure time will be determined based on the performance of the readout system and background.

\section{The Skipper-CCD Instrument}
\label{sec:instrument}
% Spelling and Grammar Checked
The DarkNESS mission integrates skipper-CCD sensors and readout electronics into a 6U CubeSat to search for DM from LEO. 
Since the first demonstration of single-electron counting with a skipper-CCD~\citep{Tiffenberg:2017aac}, the SENSEI and DAMIC-M collaborations have developed skipper-CCDs for low-mass DM detection, exploiting the sensor's sub-electron noise and fine pixelation to set world-leading 
results in the field~\citep{Crisler:2018gci, SENSEI:2019ibb, SENSEI:2020dpa, senseibeam, Sensei2023,DAMIC_2024_Mod}. Ongoing developments of the skipper-CCD technology further demonstrate the utility of sub-electron noise sensors for low-threshold rare event searches~\citep{2022oscura, OscuraSensors, CONNIE_2024}. 
The DarkNESS research program has focused on adapting skipper-CCDs and their readout electronics to the challenges of space-based operation on a NanoSatellite. Through this effort, DarkNESS will open opportunities for the efficient testing and space certification of novel imaging sensors~\citep{Botti_2024,Sofo_Haro_2024,Lapi_2024}.

\begin{figure}[t]
    \centering
    \includegraphics[width=\columnwidth]{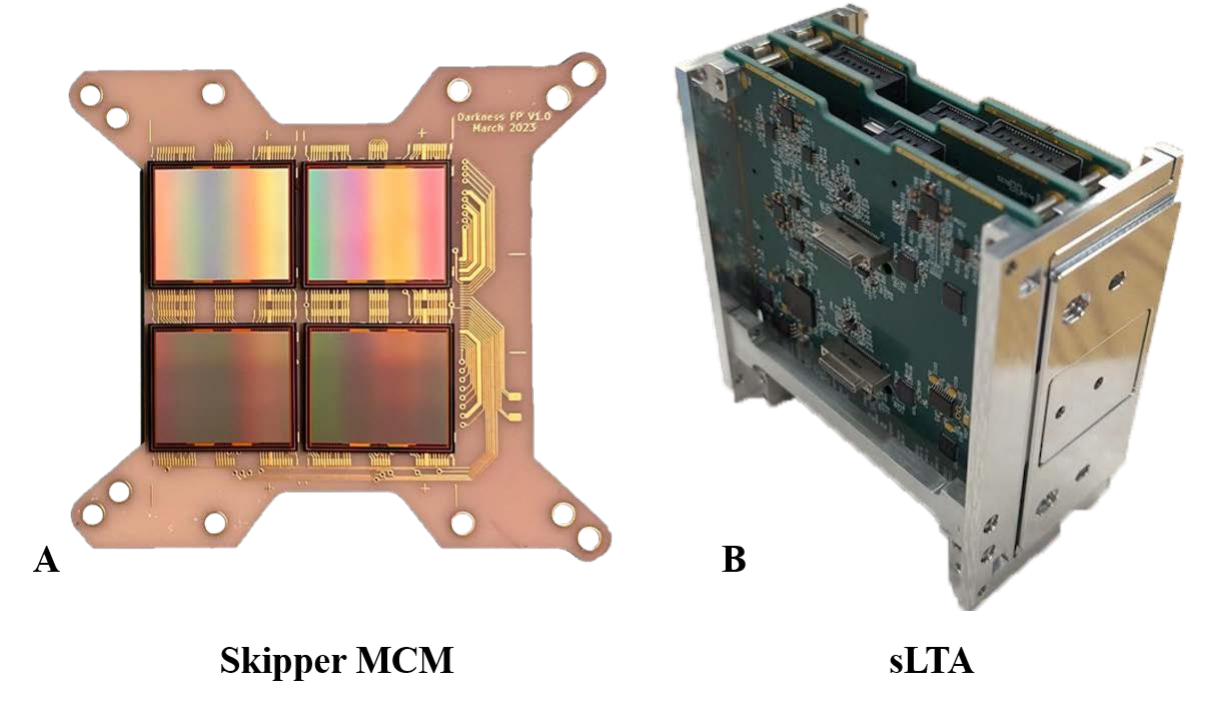}
    \caption{A) The Multi-Chip Module (MCM) designed, built, and tested for DarkNESS features the four 1.35 Mpix skipper-CCD designed by LBNL and fabricated at Microchip as part of the R\&D effort for DM experiments~\citep{OscuraSensors}. B) The space Low-Threshold Acquisition (sLTA) readout electronics housed in the DarkNESS thermal control board box~\citep{lta}.}
    \label{fig:instrument}
\end{figure}
\label{sec:payload}

\subsection{Multi-Chip Module and Readout Electronics}
The DarkNESS instrument consists four 1.3\,Mpix skipper-CCDs integrated into a Multi Chip Module (MCM), as shown in Fig.~\ref{fig:instrument}~A. 
The skipper-CCDs were designed at the Microsystems Laboratory at Lawrence Berkeley National Laboratory (LBNL), and were recently fabricated as part of a Dark Matter New Initiatives research and development effort~\citep{OscuraSensors}. 
The LBNL design uses fully-depleted detectors up to 725\,$\mu$m thick~\citep{SteveCCD2003}, providing high efficiency to soft X-rays. Specific aspects of the detectors are modified to adapt the skipper-CCD technology to X-ray astronomy. A 50\,nm thick aluminum layer is applied to the front of the silicon detector to block visible and near-IR photons, serving as an X-ray entrance window providing $>$98\% transmission of X-rays down to 1\,keV. Backside processing to thin the detectors is being explored to mitigate the effect of particle backgrounds while providing high-efficiency X-ray spectrometry up to 10\,keV.

\begin{figure}[t]
    \centering
    \begin{subfigure}[t]{0.49\columnwidth}
    \centering
    \includegraphics[width=\columnwidth]{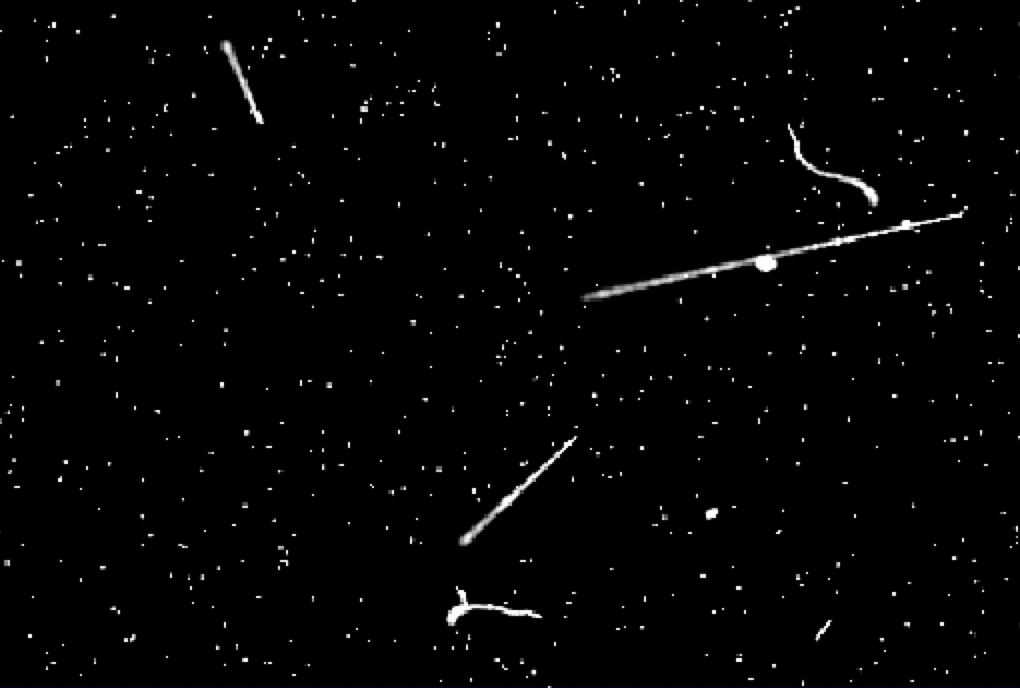}   
    \end{subfigure}
    \begin{subfigure}[t]{0.49\columnwidth}
    \centering
    \raisebox{-0.02\height}{\includegraphics[width=\columnwidth]{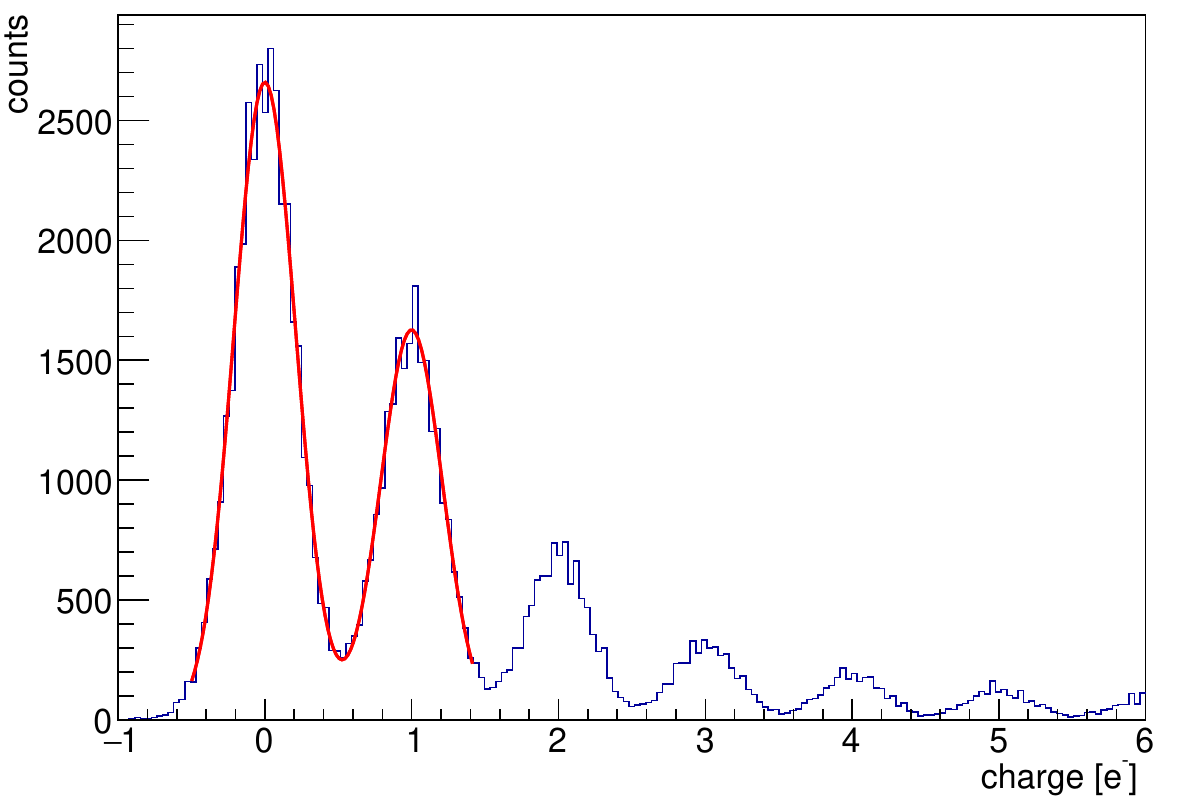}}
    \end{subfigure}
    \caption{\textbf{Left:} X-ray testing image obtained at FNAL with a prototype DarkNESS MCM operating in a vacuum chamber. The single-pixel hits represent X-ray energy depositions from a $^{55}$Fe source mounted inside the chamber. The long, straight tracks correspond to muons; other hits are likely multiple scattering electrons. 
    \textbf{Right:} Calibrated spectrum from prototype DarkNESS MCM that demonstrates the sub-electron noise counting capabilities using 300 skipper samples. The $x$-axis is in units of electrons; blue histogram shows the pixel spectrum, and the red line is a Poisson fit convolved with Gaussian noise. The individual electron peaks can be resolved with ~0.2\e of noise.}
    \label{fig:TestResults}
\end{figure}

The DarkNESS MCM uses a lightweight ceramic package to hold the four skipper-CCDs, 
shown in Fig.~\ref{fig:instrument}. A custom flex circuit is epoxied and wire bonded to the MCM, connecting the MCM to the readout electronics. A prototype MCM and flex cable has been designed and tested in a cryogenic vacuum chamber at Fermi National Accelerator Laboratory (FNAL) using a $^{55}$Fe X-ray source. Fig.~\ref{fig:TestResults} shows an image and low-energy spectra from a prototype MCM.

The readout electronics for DarkNESS are based on the Low Threshold Acquisition (LTA) system developed for skipper-CCDs~\citep{lta}. The LTA system provides the bias voltage to operate the skipper-CCD and controls the charge sequencing to read out the pixel array. To integrate the LTA into the 6U CubeSat platform, FNAL designed a compact version called the space-LTA (sLTA) shown in Fig.~\ref{fig:instrument}~B. The sLTA partitions the functionality of the LTA into three boards compatible with the PC104 standard, uses less power than the standard LTA, and includes a copper plane for improved thermal management. The sLTA requires 10\,W of power to operate the skipper-CCDs and is controlled by a NanoAvionics payload controller through an ethernet connector. The sLTA is housed in an aluminum enclosure to aid in the electronics’ thermal management in the space environment. DarkNESS X-ray observations require faster readout than direct DM applications, resulting in new requirements for the skipper-CCD front-end electronics. DarkNESS aims to achieve 
a readout speed of 250\,kPix/s per skipper sample, allowing full frame readout in 5\,s.

The skipper-CCDs are cooled to 170\,K using a Ricor K508N compact rotary cryocooler~\citep{Mok2020_K508N}, and the NanoSatellite is designed to dissipate excess heat from the sLTA and cryocooler to the radiators and described in Sec.~\ref{sec:thermal}. The instrument operating temperature requirement of 170\,K is specified to mitigate dark current due to thermal excitation of electron-hole pairs~\citep{Widenhorn_2002}. Thermal management tests are ongoing to verify the sLTA thermal behavior in simulated space environment conditions, which are described in Sec.~\ref{sec:thermal}.

\subsection{Field of View and Exposure}
\label{sec:FOV}
Given the CubeSat's volume constraints, the payload does not utilize X-ray optics. Each pixel will observe a $20^\circ$ FOV defined by four circular apertures in front of the sensor array. As shown in Fig.~\ref{fig:instrument}, the MCM provides a total collecting area of 12\,cm$^2$. DarkNESS has a notably large FOV compared with other X-ray telescopes. For example, XMM EPIC-MOS has a FOV of approximately 30~arcminutes and a collection area of 700\,cm$^2$ for 3\,keV X-rays. This means that one DarkNESS exposure observes a diffuse background flux comparable to 22 XMM EPIC-MOS images with the same exposure time. Since the skipper-CCDs can be fully depleted to a thickness of 725\,$\mu$m, increased efficiency for 10--20\,keV X-rays is expected compared to thinner CCDs used in EPIC-MOS. The low-noise skipper-CCDs provide excellent energy resolution and have demonstrated energy resolution down to the Fano limit in silicon ($\sigma_E \sim 50$\,eV at 6\,keV)~\citep{darioFanoFirst}.
DarkNESS has a technical goal of achieving Fano-limited energy resolution for 1--10\,keV X-rays in $<$1\,min readout time.
The large FOV and Fano-limited energy resolution allow DarkNESS to achieve a high sensitivity to the diffuse X-ray background.

\subsection{Skipper-CCD Considerations in LEO}
\label{LEO}
The radiation environment in LEO is expected to be a significant challenge for detector operations. To ensure the successful operation of skipper-CCDs in LEO, DarkNESS uses
Monte Carlo simulation tools~\citep{AGOSTINELLI2003250} and accelerated lifetime testing to assess the effects of operating skipper-CCDs in LEO.

\begin{figure}[t]
\includegraphics[width=0.575\columnwidth]{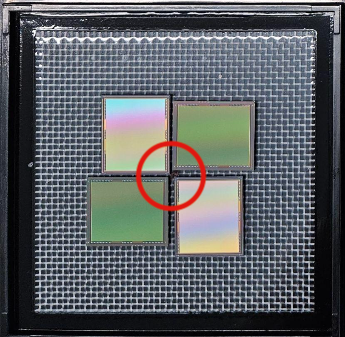}
    \includegraphics[width=0.415\columnwidth]{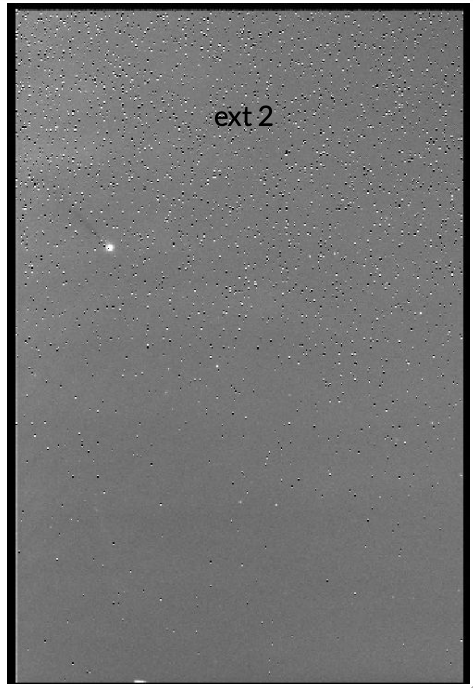}
    \centering
    \caption{\textbf{Left:}~Four skipper-CCDs in the sample holder used for proton irradiation testing. The proton beam was aligned with the center of the sample holder and covered approximately the area in the red circle. \textbf{Right:}~Image used to identify single-electron traps after irradiation using the pocket pumping technique~\citep{Janesick:2001}. The skipper-CCD shows more traps in the top portion of the image, which was exposed to the full dose of protons, compared to the bottom portion of the image that received a smaller fluence.}
    \label{fig:protonsWarrenville}
\end{figure}

To evaluate the effect of radiation damage to the skipper-CCDs throughout the mission, prototype DarkNESS skipper-CCDs were irradiated with a 217~MeV proton beam at the Northwestern Medicine Proton Center, delivering a total fluence of $1.2\times10^{10}$~protons/cm$^2$.
Despite this exposure, the skipper output stage showed no degradation, and the sensors maintained sub-electron noise performance~\citep{Roach_2024}. The robustness of the skipper amplifier enabled post-exposure studies of the density of single-electron traps in the imaging area, using the charge pumping technique described in~\cite{Janesick:2001}. These studies indicate that the protons created single-electron traps in the imaging area (see Fig.~\ref{fig:protonsWarrenville}). These defects were mainly divacancies, consistent with displacement damage observed in p-channel CCDs described in~\cite{Bebek:2002b, Hall:2017}. The radiation-induced trap density is $8\times10^4$/cm$^2$, corresponding to 0.18 traps per pixel, given the $15\times15~\mu$m$^2$ pixel size. This exposure was equivalent to a fluence of $3.4\times10^9$~protons/cm$^2$ for 12.5~MeV protons, roughly four times the total fluence expected over one year of DarkNESS operations. 

At a typical LEO altitude of 450\,km, the expected trapped proton fluence is $9\times10^8$~protons/cm$^2$ at 10~MeV\footnote{\url{https://www.spenvis.oma.be}}. Based on these results, we do not expect any degradation of the skipper amplifier throughout the DarkNESS mission. Still, we note that approximately 5\% of the CCD pixels will accumulate single-electron traps per year in LEO. These traps can degrade the CCD's charge transfer efficiency and require masking techniques around high-energy events to eliminate spurious low-energy events during low-mass DM searches~\citep{SENSEI:2020dpa}. These recent results of the irradiation test corroborate previous work with similar CCDs~\citep{RadToleranceLBNL}, which indicated that the LBNL CCDs perform well after exposure to space radiation.

Another challenge for operating skipper-CCDs in LEO is the generation of very low-energy hits (a few electrons) produced by ionizing radiation in the detector, which was discussed recently in~\cite{Gaido_2024}.
As highly-energetic charged particles traverse the detector, Cherenkov photons are produced that comprise a background for low-threshold direct DM searches~\citep{SecondaryRad_Rouven}. 
To mitigate this effect, DarkNESS will use an imaging analysis framework with tunable selection criteria that can define a region around high-energy events to be removed from the low-energy analysis (as done for the DM search in~\cite{SENSEI:2020dpa}).

\section{The DarkNESS Observatory}
\label{sec:design}

The DarkNESS observatory integrates an actively chilled skipper-CCD payload within a NanoAvionics M6P 6U platform. Subsystem constraints govern the architecture, including attitude configurations, radiative heat rejection, and orbit-wise power balance. Thermal dissipation occurs through body-mounted radiators, with performance set by Earth and deep-space view factors. The attitude control strategy incorporates secondary pointing constraints that manage radiative efficiency under changing observational conditions. These architectural elements, refined through orbit analysis and thermal hardware integration testing at UIUC, support sustained low-temperature operation and science data acquisition under varied orbit considerations.

\subsection{Concept of Operations}

\begin{figure*}[ht]
\includegraphics[width=0.9\textwidth]{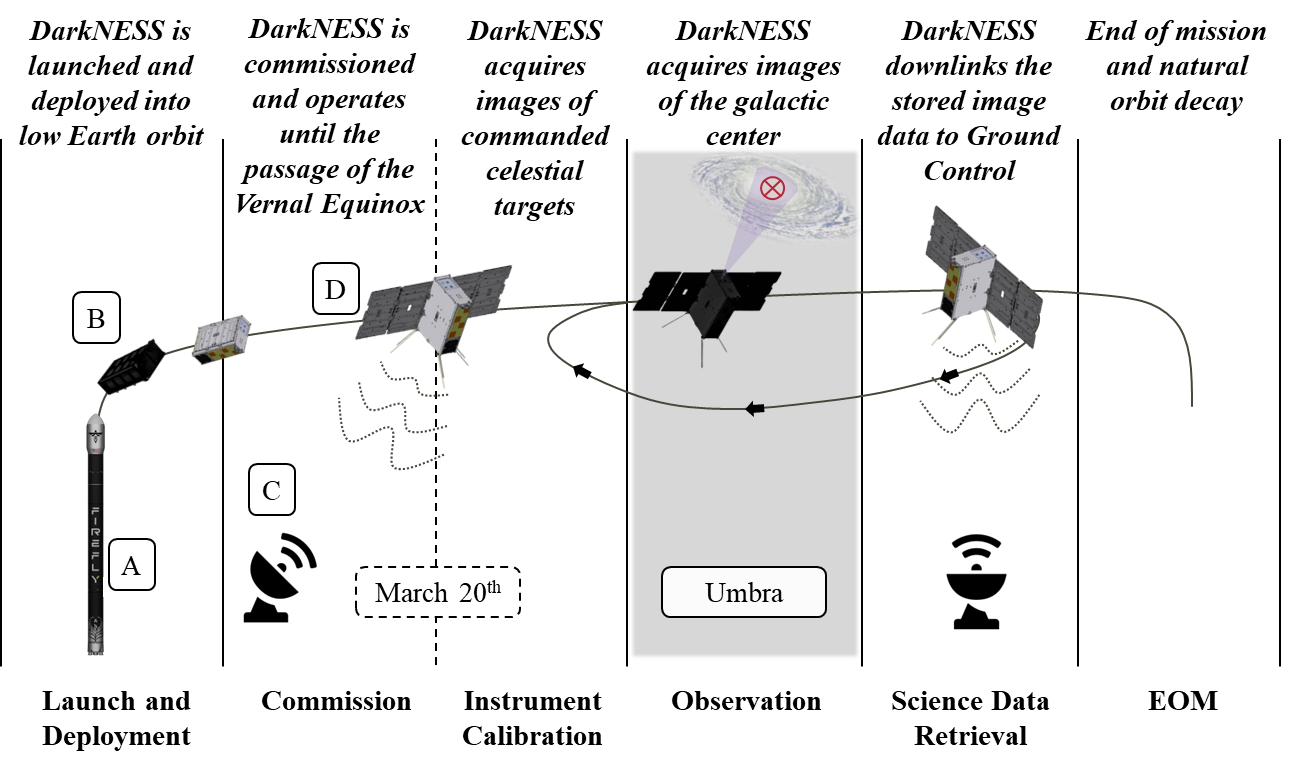}
\centering
\caption{\textbf{Concept of Operations:}. The Firefly Alpha launch vehicle (A) delivers DarkNESS to LEO and is deployed by the Exolaunch NOVA dispenser (B). The ground station (C) commands DarkNESS (D) subsystems commissioning. After the passage of the Vernal Equinox, DarkNESS begins instrument calibration and performs repeated observations towards Sagittarius and Cygnus during umbral passages. Image histograms and raw files are transmitted between observation sessions. The mission concludes with natural orbit decay within five years.}
\label{fig:conops}
\end{figure*}

The DarkNESS concept of operations (Fig.~\ref{fig:conops}) is structured to meet its scientific objectives within the volume, power, and pointing constraints of a 6U NanoSatellite while maintaining deployment orbit flexibility. The observatory, housed in an Exolaunch NOVA dispenser, will launch aboard Firefly Aerospace’s Alpha vehicle. After deployment, the mission enters commissioning, entailing subsystem checkouts and payload configuration, before transitioning into science operations.

The science phase is defined by observations targeting the Galactic Center (towards Sagittarius) between March 20 and September 23, 2026, spanning 187 days between the Vernal and Autumnal Equinoxes. This period offers visibility during umbral passages, reducing solar background and supporting stable thermal conditions. Pointing constraints are set by inertial targets and secondary considerations such as radiator exposure to deep space and view factors to Earth. Observations targeting the solar apex (towards Cygnus) are viable throughout the year, as described in Sec.~\ref{sec:orbit}. The mission concludes with natural orbit decay, abiding by space debris mitigation guidelines~\citep{fcc2024orbital_debris}.

Ground operations are conducted in collaboration with Illinois State University’s (ISU) NanoSatellite Ground Station. UIUC commands the observatory via UHF from the ISU station, and science data are downlinked via S-band during scheduled passes. Data are then transferred to FNAL for processing and archiving.

\subsection{Observatory Configuration}
\label{sec:configuration}

\begin{figure}[h]
\includegraphics[width=0.9\columnwidth]{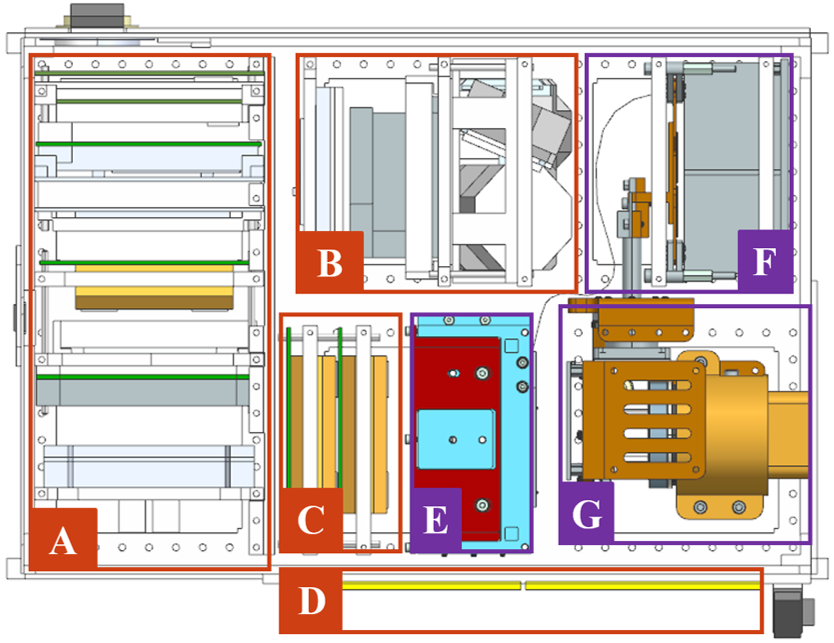}
    \centering
    \caption{\textbf{Internal Configuration:} major components include the avionics stack (A) includes the flight computer, payload controller, and dual S-band radios. Attitude control components (B) include reaction wheels, an inertial measurement unit (IMU), and magnetorquers. The electrical power system (C) includes the batteries and power distribution board. Two S-band patch antennas (D) are mounted to the $-Z$ face. The payload sLTA electronics (E) control skipper-CCD readout and connect to the MCM (F) via a flexible printed circuit. The cryocooler (G) maintains the MCM at 170\,K through a fixed-point thermal control loop.}
    \label{fig:subsystems}
\end{figure}

DarkNESS will be deployed via Exolaunch’s NOVA 6U/8U dispenser, which accommodates up to 25\,mm external protrusion—sufficient for the body-mounted radiators and stowed dual-deployable solar arrays. The design supports flexible accommodation into a range of LEO deployments, including $\sim$500\,km altitude Sun-Synchronous Orbits (SSO), and ISS-like mid-inclination orbits. Deployment is scheduled no earlier than mid-2026, with the final orbit determined at manifest.

\begin{figure}[t]
    \centering \includegraphics[width=0.9\columnwidth]{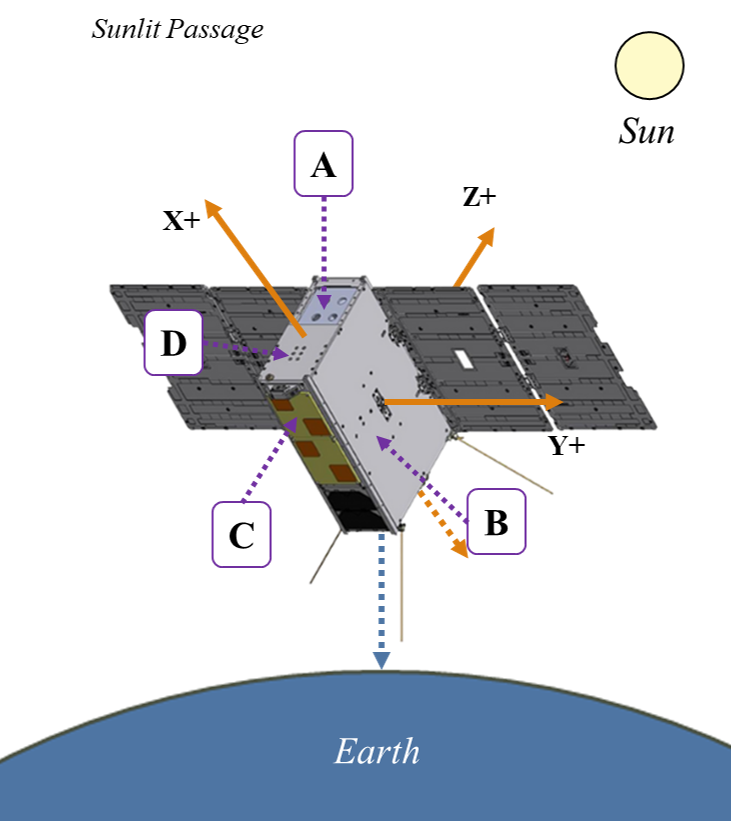}
    \caption{\textbf{External Configuration:} The window aperture (A) sets the payload MCM field-of-view. Radiators on the $\pm Y$ faces (B) and the $+X$ face (D) support passive thermal control, including dissipation of cryocooler waste heat and payload sLTA. Redundant S-band patch antennas (C) provide telemetry and science data downlink. In Sunlit mode, the $+Z$ face tracks the Sun while the $-X$ face aligns toward nadir to balance the radiator view factors of Earth.}
    \label{fig:sunlit}
\end{figure}

The observatory subsystem configuration (Fig.~\ref{fig:subsystems}) is driven by the thermal requirements of the MCM, which operates at 170~$\pm$~5\,K. The cryocooler maintains this temperature through fixed-point closed-loop control, resulting in a continuous power demand. Waste heat from the cryocooler and sLTA is passively rejected using copper straps connected to three body-mounted radiator panels, oriented to maintain exposure to deep space while balancing the view factor to Earth, particularly during umbral passages. Three external surfaces are allocated to radiators to preserve uninterrupted radiator exposure to deep space, constraining solar array placement and aperture alignment. The external configuration of DarkNESS is illustrated in Fig.~\ref{fig:sunlit}. Power requirements during umbra determine the sizing of a 2S7P battery and dual-deployable solar arrays. These arrays recharge the battery over one Sunlit passage, maintaining balance across orbit-wise science duty cycles.

\begin{figure}[t]
    \centering    
    \includegraphics[width=0.9\columnwidth]{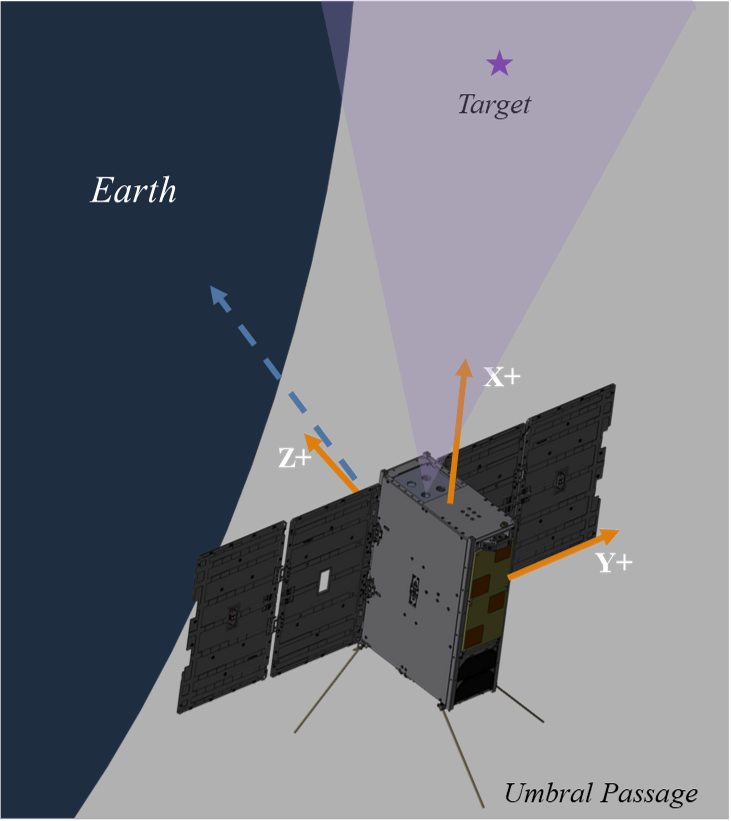}
    \caption{\textbf{Inertial Science Pointing:} Observation targeting determines the primary attitude constraint and aligns the $+X$ axis to the objective (Sagittarius A$^*$ or Cygnus X-1) during umbral passage. The $+Z$ axis aligns toward nadir to shield radiator panels from exposure to the Earth using the deployed solar array. Earth and Moon often obstruct portions of the MCM field-of-view}
    \label{fig:umbra}
\end{figure}

Attitude operations are divided into three modes: Sun-tracking (Fig~\ref{fig:sunlit}), Ground Station Pointing, and Inertial Science Pointing (Fig.~\ref{fig:umbra}). Each includes a primary axis constraint for mission function and a secondary constraint to orient radiators toward deep space or away from Earth and Sun. 

\begin{itemize}
    \item \textbf{Sun-Tracking:} Primary axis (+Z) aligns with the Sun. Secondary axis (-X) points nadir to improve radiator view to deep space and maintain detector window orientation away from Earth.

    \item \textbf{Ground Station Pass:} Primary axis (-Z) aligns with the S-band antenna boresight. Secondary (+X) points nadir or along-track to reduce Earth-view of radiators.
    
    \item \textbf{Inertial Science Pointing:} Primary axis (+X) aligns with the science target. Secondary axis (+Z) points nadir, leveraging solar arrays to block Earth IR from radiator view
\end{itemize}

The ADCS system supports coarse inertial pointing with a $4.5^\circ$ ($3\sigma$) absolute error and $<1.5^\circ$ ($3\sigma$) knowledge error using fused measurements from the magnetometer, Sun sensors, gyroscope, and GPS. A 7-state Extended Kalman Filter generates attitude estimates for a software PID controller. Six reaction wheels are configured to maintain three-axis control and compensate internal secular torques from the cryocooler motor and compressor assemblies. 

The communication system includes two S-band radios. The SatLab SRS-4 supports high-rate science downlink (up to 3\, Mbps) for image histograms (2.5\,kB) and raw image (32\,MB) files. The SRS-3 provides a redundant S-band path for telemetry and image histograms. A KNA UHF transceiver handles telemetry and command uplink at 401–402\,MHz (9.6\,kbps max). All payload communication is routed through a dedicated controller interfaced with the sLTA via 1000BASE-T Ethernet.

\section{Orbit Considerations}
\label{sec:orbit}

The two science objectives outlined in Section~\ref{sec:science}—probing strongly interacting DM and X-rays from decaying DM—require repeated observations of two distinct regions on the celestial sphere: the Galactic Center (Sagittarius) and Cygnus (Table~\ref{tab:targets}). These coordinates serve as reference directions for broader regions of interest rather than point-source targets. For decaying DM searches, the instrument integrates faint X-ray flux across a wide solid angle ($20^\circ$) centered near Sagittarius A$^*$, observing the Galactic plane where the expected signal rate of photons due to DM decay is highest due to the high D-factor~\cite{Evans_2016}.

Orbit dynamics introduce time-dependent Earth obstruction of both targets during DarkNESS's passage through umbra (pointing configuration depicted in Fig.~\ref{fig:umbra}). To evaluate target visibility windows under these conditions, along with geometric intrusions from the Moon within the FOV, an obstruction analysis adopts a $40^\circ$ keep-out zone to account for off-axis response and the extended exposure of edge pixels within the CCD array (Fig.~\ref{fig:instrument}). Unlike Sagittarius targeting, observations towards Cygnus utilize Earth obstruction to search for modulation signatures of strongly interacting DM. This section outlines the obstruction modeling approach, observation feasibility across representative orbit domains, and implications for data acquisition strategies.

%----------------------- Observation Targets Table -----------------------
\begin{table}[h]
    \centering
    \caption{Observation Targets and ICRF Coordinates}
    \label{tab:targets}
    \renewcommand{\arraystretch}{1.2}
    \setlength{\tabcolsep}{6pt}
    \resizebox{\columnwidth}{!}{%
    \begin{tabular}{lccc}
        \hline
        \textbf{Target} & \textbf{RA} & \textbf{Dec} & \textbf{Science Objective and Observation Plan} \\
        \hline
        \multicolumn{4}{@{}l}{\textit{Science Phase: Vernal–Autumnal Equinox}} \\
        Sagittarius A* & 17h45m40s & $-29^\circ00^\prime28^{\prime\prime}$ & X-ray decay search; 600×15-min exposures \\
        \multicolumn{4}{@{}l}{\textit{Science Phase: Full Year}} \\
        Cygnus X-1     & 19h58m21.7s & $+35^\circ12^\prime06^{\prime\prime}$ & Modulation study; 450×10-min exposures \\
        \hline
    \end{tabular}
    }
    \caption*{\footnotesize VLBI-based ICRF coordinates: Sagittarius A* (towards Galactic Center)~\citep{Gordon2023:sagA*}; Cygnus X-1 (near Solar Apex, DM wind direction)~\citep{Stirling2001:CygnusX1}.}

\end{table}

\subsection{Celestial and Geometric Constraints}

\begin{figure}[h] \centering \includegraphics[width=0.8\columnwidth]{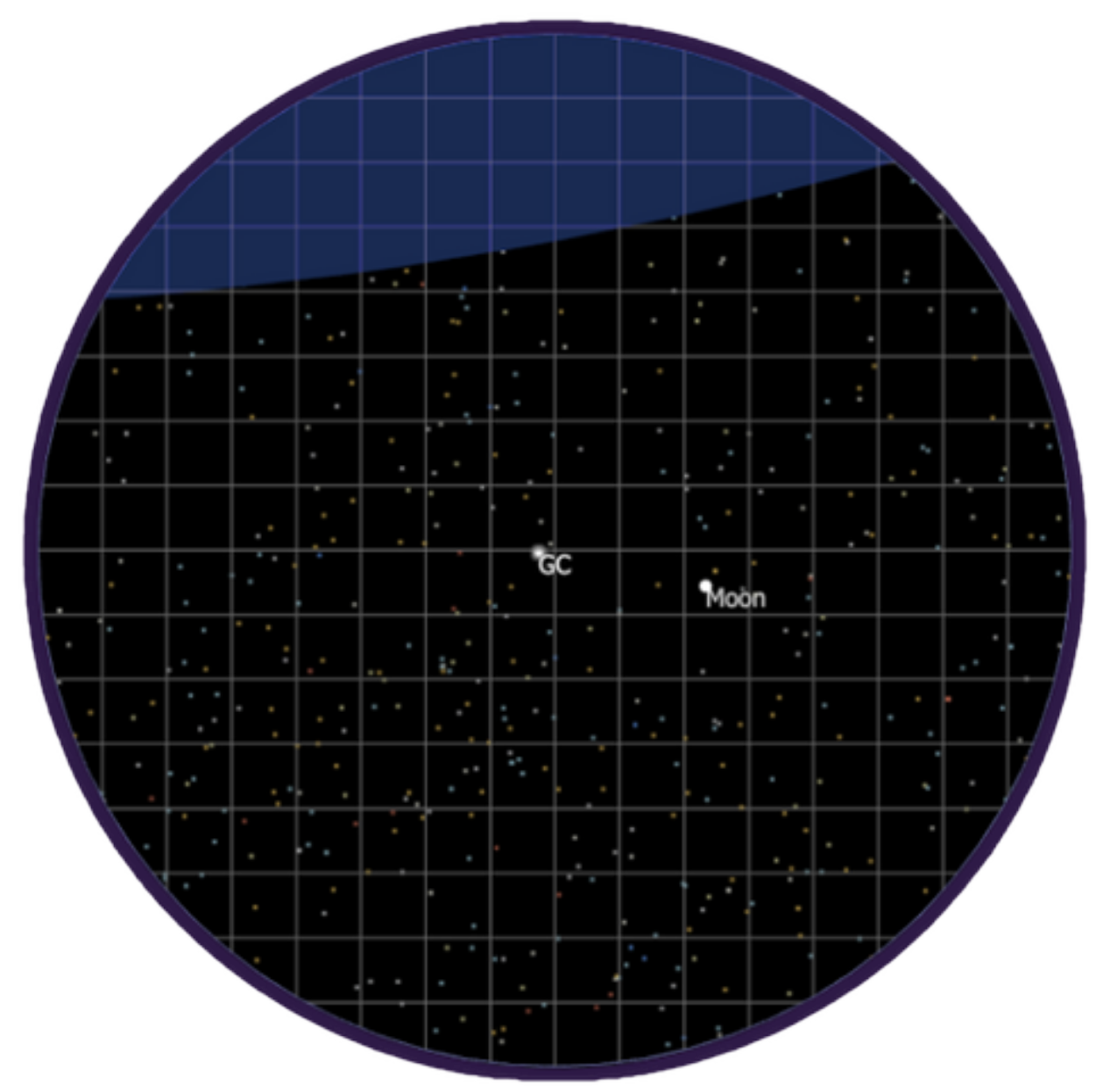} \caption{Simulated MCM with a $40^\circ$ FOV using FreeFlyer (a.i. Solutions). Earth and Moon obstruct the target direction during an observation attempt in April 2026. This obstruction mapping constrains acquisition planning by identifying the timing and extent of geometric visibility used to quantify observation windows.} \label{fig:simulated_sensor} \end{figure}

The Galactic Center lies near the anti-solar direction during Northern Hemisphere summer, aligning it with umbral passages between the Vernal and Autumnal Equinoxes. This geometry supports extended exposures for faint X-ray detection by reducing solar background and mitigating radiative loading on the MCM. Observations targeting the Galactic Center prioritize unobstructed windows clear from Earth and lunar intrusion. Figure~\ref{fig:simulated_sensor} shows a simulated observation from early Spring 2026 affected by combined obstruction.

As Earth progresses in its heliocentric orbit, the umbral axis traces a full rotation in inertial space. Concurrently, the Right Ascension of the Ascending Node (RAAN) of the observatory’s orbit regresses due to Earth’s oblateness, modeled here using a spherical harmonic representation of the geopotential~\cite{vallado2022ch9,prussing2012Ch11,Kaula1966:geodesy}. This regression rotates the orbit plane within the geocentric celestial frame. These motions shift the alignment between the observatory's boresight, umbral axis, and inertial targets throughout the mission. The resulting viewing geometry imparts Earth obstruction and reshapes the frequency and duration of open observation windows. The effect is most pronounced in mid-inclination LEO, where seasonal and orbital motion compound to vary the Sun’s angle of incidence on the orbit. In contrast, a Sun-synchronous orbit (SSO) matches nodal regression to Earth's heliocentric motion, maintaining a fixed orbit–Sun geometry in the rotating Sun-relative frame. This removes seasonal variation in the orbit plane alignment with umbra and reduces time-dependent Earth obstruction. A Local Time of Ascending Node (LTAN) of Noon favors consistent anti-solar pointing during umbral passage, improving access to the Galactic Center between the Vernal and Autumnal Equinoxes. The RAAN regression produces a slow drift in the orbit–Sun geometry, driving long-term variation in umbral duration throughout the mission~\citep{srm2019:LEOumbratime}.

The DM wind remains fixed in the Galactic frame, but its apparent direction relative to the orbit evolves with season. In particular, the angular separation between the umbral axis and the Cygnus vector narrows near the June solstice, favoring unobstructed observations, and widens near December, increasing the occurrence of complete shadowing of the MCM FOV.

Observations toward Cygnus are used to study directional modulation from strongly interacting sub-GeV DM. When Earth enters the Cygnus line-of-sight during umbral passages, the MCM experiences geometric shadowing—defined as a reduction in the accessible DM flux due to angular occlusion~\citep{emken:JCAP2019}. Each passage may contain transitions between unobstructed, partially obstructed, and fully shadowed regimes. These conditions define the observation windows for both science objectives and are mapped into the mission timeline to guide scheduling and data collection.

\subsection{Obstruction Regimes}

Target observations during umbral passages are evaluated using FreeFlyer (a.i. Solutions) and a Python-based simulation. The tool models the MCM’s conical FOV and determines the time and duration of intrusions from Earth and Moon. Each passage is assessed for obstruction regimes suitable for science acquisition for both objectives.

Observations toward Sagittarius prioritize 15-minute unobstructed windows to mitigate background contributions and maintain the thermal stability needed for low-noise imaging (see Section~\ref{sec:instrument}).
Conversely, observations toward Cygnus interpret the distinct obstruction regimes associated with geometric modulation studies (see Section~\ref{sec:science}). The MCM maintains full angular access to the incoming DM flux when Cygnus remains unobstructed. Partial Earth obstruction attenuates this access by reducing the MCM's solid angle of exposure. In completely obstructed cases, Earth blocks the Cygnus direction entirely, resulting in geometric shadowing as modeled in~\cite{emken:JCAP2019}. These modulated regimes—unobstructed, partially obstructed (attenuated), and completely obstructed (shadowed)—form the segmentation basis for useful observation windows interpreted from the obstruction analysis.

\subsection{Minimum Science Success Criteria}

To quantify mission success, DarkNESS defines minimum observation thresholds derived from the sensitivity requirements outlined in Section~\ref{sec:science}. These thresholds translate detector mass, exposure time, and geometric access into a minimum number of valid observations based on conditions set in the obstruction analysis framework.

For the strongly interacting DM search, a 0.1 gram-month exposure toward Cygnus is required to achieve the projected discovery reach. This corresponds to 450 umbral observations, each 10 minutes in duration and split between unobstructed and shadowed regimes.

For the decaying DM search, 25 hours of clear, unobstructed 15-minute exposures toward the Galactic Center are required to reach the projected sensitivity. The mission baseline of 600 unobstructed observations provides 150 hours of exposure time. Table~\ref{tab:success_criteria} summarizes the minimum number of observations and data products required for meaningful scientific analysis and archival value. 
%----------------------- Minimum Success Criteria Table -----------------------
\begin{table}[h]
\raggedright
\caption{Minimum Success Criteria for Scientific Data Collection}
\label{tab:success_criteria}
\renewcommand{\arraystretch}{1.0}
\setlength{\tabcolsep}{4pt}
\resizebox{\columnwidth}{!}{%
{\small
\begin{tabular}{@{}lccc@{}}
    \hline
    \textbf{Target} & \textbf{No. Obs.} & \textbf{Histograms} & \textbf{Images} \\
                    &                   & (per Obs.)              & (10\% of Obs.)  \\
    \hline
    Sagittarius (Unobstructed) & 600 & 600 & 60 \\
    Cygnus (Shadowed)       & 225 & 225 & 25 \\
    Cygnus (Unobstructed)          & 225 & 225 & 25 \\
    \hline
    \textbf{Science Data:} & 1100 & 1100 & 110 \\
    \textbf{Data to Downlink:} & & $\sim$2.8 MB & $\sim$3.6 GB \\
    
\end{tabular}
}
}
\end{table}

\subsection{Representative Orbit Analysis}

Firefly Aerospace’s DREAM 2.0 program awarded a rideshare launch opportunity for DarkNESS to the student-led team at the University of Illinois. Because the final orbit will be assigned at manifest, the mission design accommodates a range of deployment scenarios typical of commercial LEO opportunities. To bound the trade space, the analysis evaluates two representative domains: (1) a mid-inclination orbit with a semi-major axis of 6788 km, consistent with prior ISS deployments, and (2) a SSO with a Local Time of Ascending Node (LTAN) of Noon and a semi-major axis of 6888 km.

These domains define a geometric envelope for evaluating the obstruction landscape over the science season. Earth and lunar intrusions into the MCM FOV are mapped using an obstruction model. The mid-inclination case is simulated across a span of Day-of-Flight (DoF) RAAN values from $0^\circ$ to $330^\circ$ in $30^\circ$ increments. The SSO case is fixed by the choice of LTAN (Noon). The analysis quantifies the dates and durations of obstruction events during observations targeting Sagittarius and Cygnus for each orbit domain.

\begin{figure}[h]
\centering
\includegraphics[width=\columnwidth]{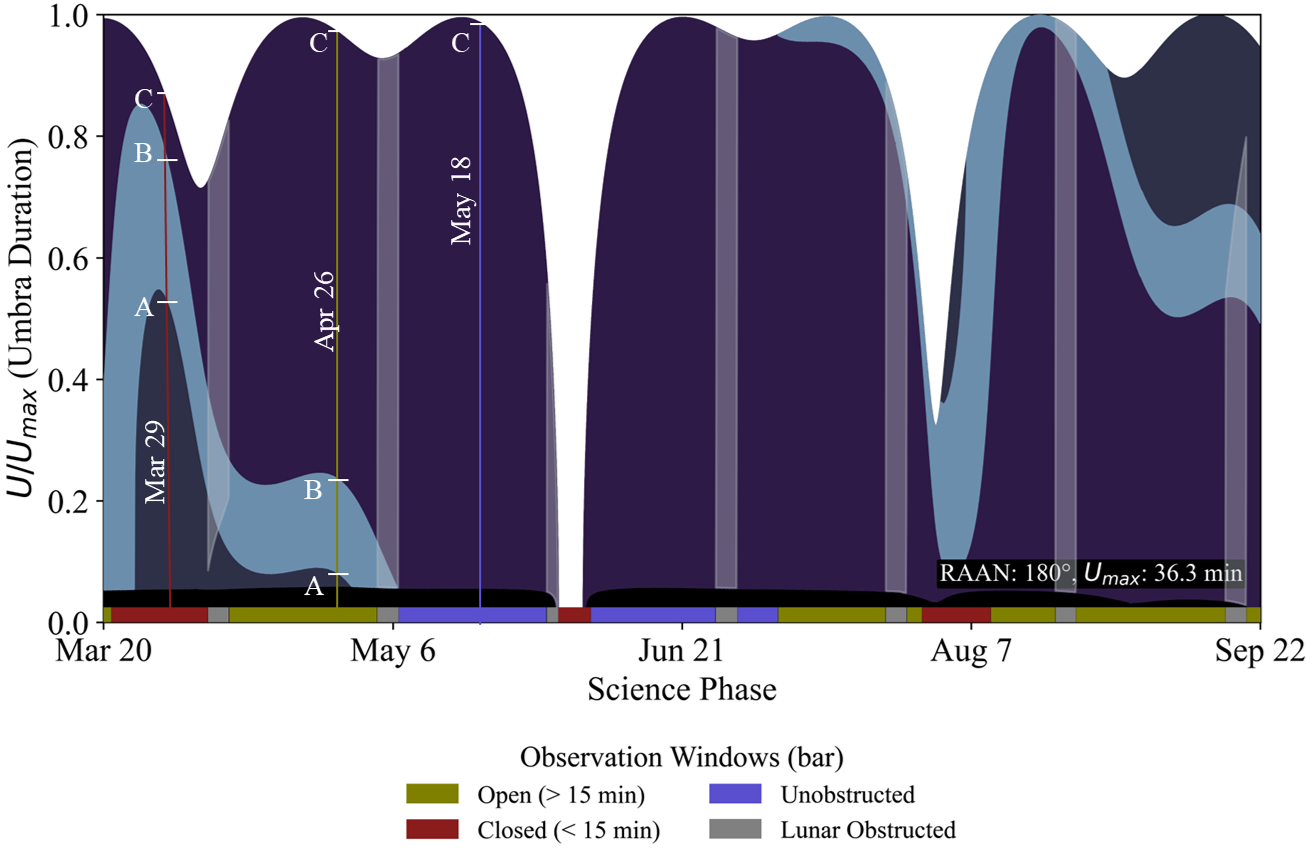}
\caption{\textbf{Sagittarius:} The simulated observation landscape targeting decaying DM X-ray signatures for an ISS-like orbit domain with a DoF RAAN of 180$^\circ$. The vertical axis represents the fractional progression through each umbral passage, normalized by $U_\mathrm{max}$, while the horizontal axis spans the science phase targeting Sagittarius. Shaded regions indicate geometric visibility: unobstructed target (Deep Iris), partial lunar obstruction (Gray), partial Earth obstruction with the target in view (Mid-Blue), and complete Earth obstruction (Deep Blue). A Boolean bar identifies passages with an unobstructed target duration exceeding 15 minutes (Olive). For example, on April 26, the FOV is completely obstructed at the beginning of the passage (A), the target becomes visible after Earth obstruction subsides (B), and remains in view until the end of the passage (C). Observation opportunities shift with RAAN.}
\label{fig:landscapeRAAN180}
\end{figure}

Figure~\ref{fig:landscapeRAAN180} shows the obstruction landscape for a representative mid-inclination deployment with a RAAN of $180^\circ$ near the Vernal Equinox, which marks the beginning of the science phase. Results across all RAAN values are summarized in Tables~\ref{tab:GCsummary}~and~\ref{tab:CYGNUSsummary}, and visualized in the seasonal obstruction landscape figures (e.g., Fig.~\ref{fig:GCdomains}).

% Landscape Figures

\subsubsection{Observations Targeting Sagittarius}

The Noon LTAN SSO maintains a stable alignment with the anti-solar axis due to its synchronized precession with Earth’s heliocentric orbit. This orientation reduces seasonal and RAAN-dependent obstruction during umbral passages. Simulations yield 1568 unobstructed windows for objectives targeting the Galactic Center, providing a margin against the science requirement of 600 unobstructed observations (Table~\ref{tab:GCsummary}).

The mid-inclination domain shows greater variation across the RAAN span. Simulations average approximately 945 unobstructed windows, with increased obstruction during late spring and mid-summer. Open windows (over 1100 on average) contain a 15-minute unobstructed segment emerging within the passage. These windows remain scientifically useful if observations are scheduled to avoid obstructed passage segments. Windows without a 15-minute unobstructed segment emerging within the passage are closed to the science objective targeting the Galactic Center.

\begin{figure}[t]
\centering
\includegraphics[width=\columnwidth]{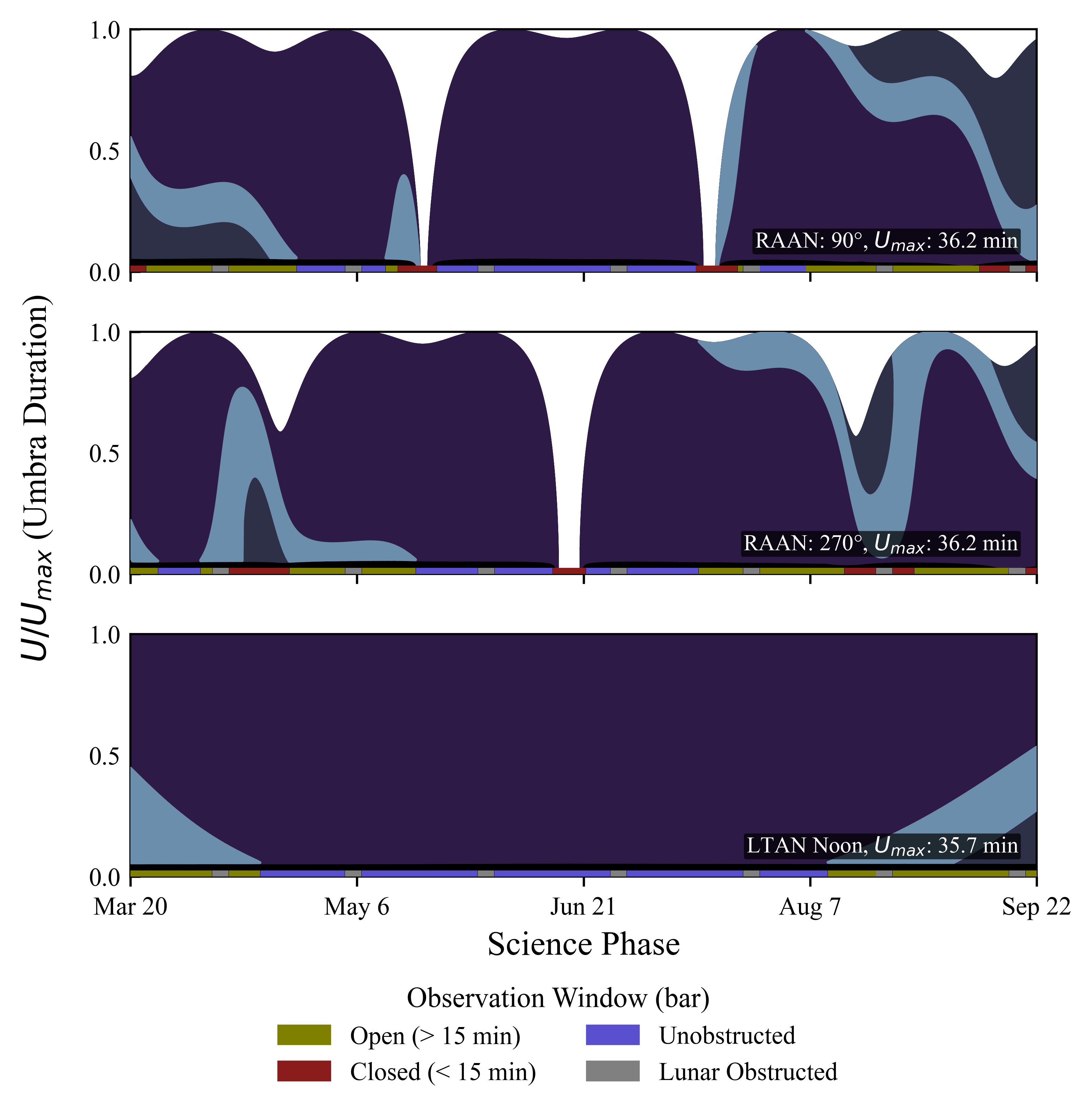}
\caption{\textbf{Sagittarius:} The simulated obstruction landscape targeting decaying DM X-ray signatures during the 2026 Vernal-Autumnal Equinoxes. The landscape shifts with instances for the mid-inclination ISS-like orbit with DoF RAAN of $90^\circ$ and $270^\circ$ and the SSO with Noon LTAN simulated in the obstruction analysis. The Galactic Center (Deep Iris) undergoes periods of partial obstruction (Mid-Blue) and full obstruction (Deep-Blue), primarily in early to mid-spring and early autumn, while the summer months offer extended clear observation windows. The Boolean bar below each plot designates observation windows. Open (Olive) and unobstructed (Iris) indicate feasible observations, with a 15-minute window emerging in partially obstructed umbral passages. For clarity, lunar obstruction (Gray) is omitted from the landscape but is represented in the Boolean bar.}
\label{fig:GCdomains}
\end{figure}

Figure~\ref{fig:GCdomains} overlays the observation windows targeting Sagittarius during the Vernal to Autumnal Equinoxes. Both orbit domains satisfy the minimum observation criteria for the decaying DM search, with reserve windows available for scheduling margin or reallocation. Observations conducted from a Noon LTAN SSO orbit provide the benefit of simplified operations planning, while mid-inclination deployments require a scheduling effort to avoid fragmented window segments, penumbral conditions, and periods of increased radiative loading on the MCM during partially obstructed passages. Reserve windows may also be allocated to the modulation objectives targeting Cygnus.

%----------------------- Galactic Center Summary Table -----------------------
\begin{table}[t]
\raggedright
\caption{Simulated observation windows targeting Sagittarius (2026 Vernal - Autumnal Equinoxes)}
\label{tab:GCsummary}
\renewcommand{\arraystretch}{1.0}
\setlength{\tabcolsep}{4pt}
\resizebox{\columnwidth}{!}{%
{\small
\begin{tabular}{@{}lccccc@{}}
    \hline
    \textbf{Orbit Domain} & \textit{Windows:} & \textbf{Unobstructed} & \textbf{Lunar} & \textbf{Open} & \textbf{Closed} \\
    \hline
    Mid-incl. RAAN$^*$   &  & 945  & 365  & 1132  & 470\\
    SSO LTAN Noon     &  & 1568 & 363  & 904   & 0\\
    \hline
\end{tabular}
}
}
\caption*{\footnotesize * Mid-inclination domain is summarized from simulations spanning RAAN from $0^\circ$ to $330^\circ$ in $30^\circ$ increments with the mean windows reported}
\end{table}

\subsubsection{Observations Targeting Cygnus}

Cygnus observations are structured around Earth-induced modulation, which contrasts with the unobstructed windows prioritized for Sagittarius. Observation regimes are summarized in Table~\ref{tab:CYGNUSsummary} and categorized as unobstructed, attenuated, or shadowed based on the degree of Earth intrusion into the observatory’s conical FOV. These geometric regimes support a directional modulation search for strongly interacting DM, with shadowed and attenuated conditions offering sensitivity to angular flux variation. 

%----------------------- Cygnus Summary Table -----------------------
\begin{table}[h]
\raggedright
\caption{Simulated observation regimes targeting Cygnus (full-year)}
\label{tab:CYGNUSsummary}
\renewcommand{\arraystretch}{1.0}
\setlength{\tabcolsep}{4pt}
\resizebox{\columnwidth}{!}{%
{\small
\begin{tabular}{@{}lccccc@{}}
    \hline
    \textbf{Orbit Domain} & \textit{Regimes:} & \textbf{} & \textbf{Unobstructed} & \textbf{Attenuated} & \textbf{Shadowed} \\
    \hline
    \multicolumn{6}{@{}l}{\textit{2026 Vernal to Autumnal Equinox}}\\
    % \multicolumn{6}{@{}l}{\textbf{\textit{Cygnus}}}\\
    Mid-incl. RAAN 90°  &  &   & 655    & 767  & 1416\\
    Mid-incl. RAAN 270° &  &   & 533    & 846  & 1461\\
    SSO LTAN Noon     &  &     & 0      & 494  & 2341\\
    % \multicolumn{6}{@{}l}{\textbf{\textit{Cygnus $20^\circ$ FOV}}}\\
    % Mid-incl. RAAN 90°  &  &   & 1416    & 264  & 1158\\
    \hline
    \multicolumn{6}{@{}l}{\textit{2026 Autumnal to 2027 Vernal Equinox}}\\
    % \multicolumn{6}{@{}l}{\textbf{\textit{Cygnus}}}\\
    Mid-incl. RAAN 90°  &  &   & 127    & 518  & 2113\\
    Mid-incl. RAAN 270° &  &   & 0    & 416  & 2349\\
    SSO LTAN Noon     &  &     & 0      & 393  & 2141\\
    % \multicolumn{6}{@{}l}{\textbf{\textit{Cygnus $20^\circ$ FOV}}}\\
    % Mid-incl. RAAN 90°  &  &   & 405    & 240  & 2113\\
    \hline
\end{tabular}
}
}
\end{table}

\begin{figure}[t]
\centering
\includegraphics[width=\columnwidth]{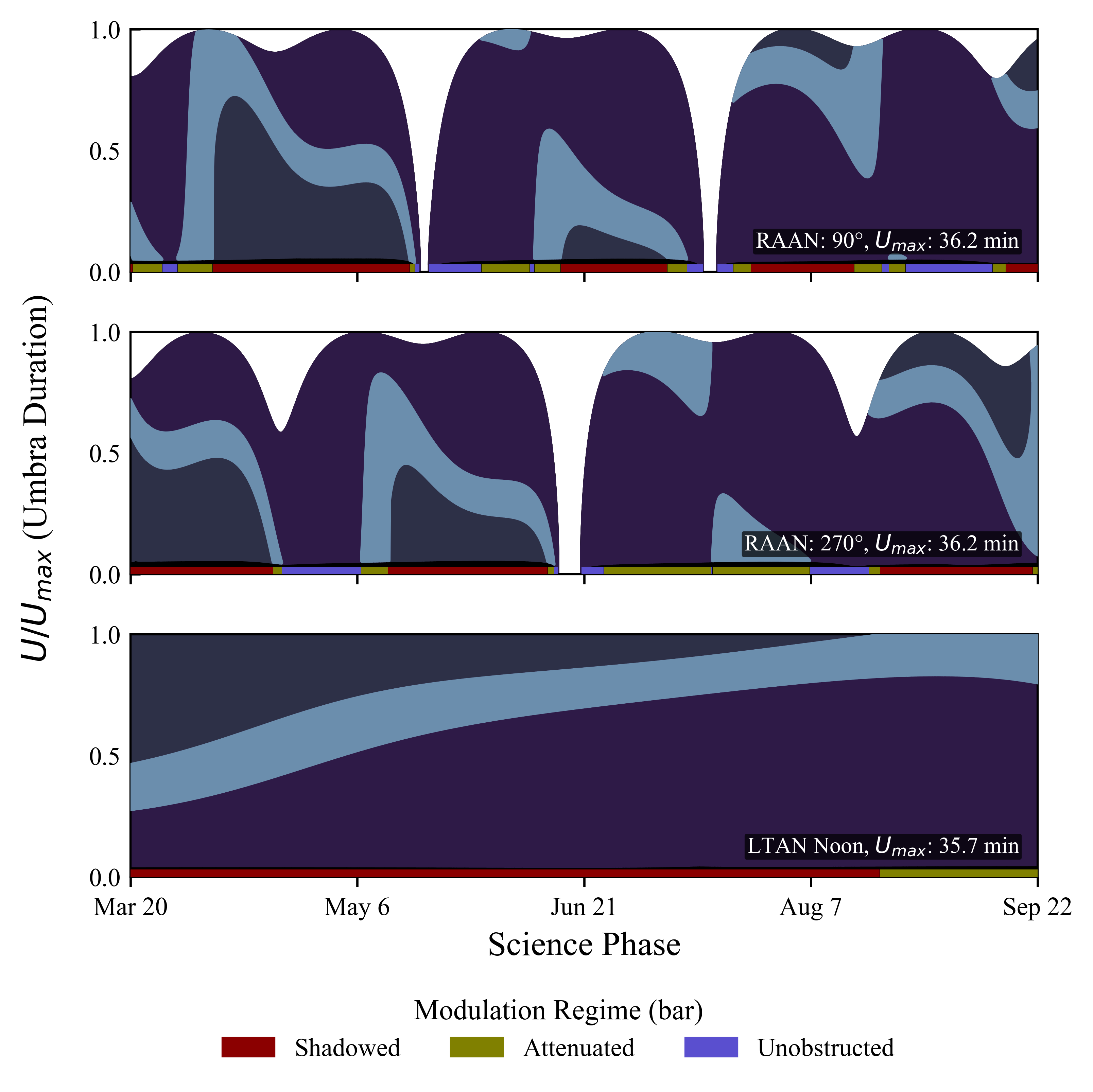}
\caption{\textbf{Cygnus:} The simulated obstruction landscape targeting strongly interacting DM flux and modulated studies during the 2026 Vernal–Autumnal Equinoxes, comparing RAAN values of $90^\circ$ and $270^\circ$ for mid-inclination deployments. Continuous shading follows the same gradient convention defined in Fig.~\ref{fig:GCdomains}. Observation regimes are classified by the Boolean bar (Crimson: shadowed, Olive: attenuated, Iris: unobstructed). Seasonal umbra rotation with respect to Cygnus fragments access to unobstructed windows and supports segmentation of modulation conditions.}
\label{fig:VACYGNUSdomains}
\end{figure}

\begin{figure}[h]
\centering
\includegraphics[width=\columnwidth]{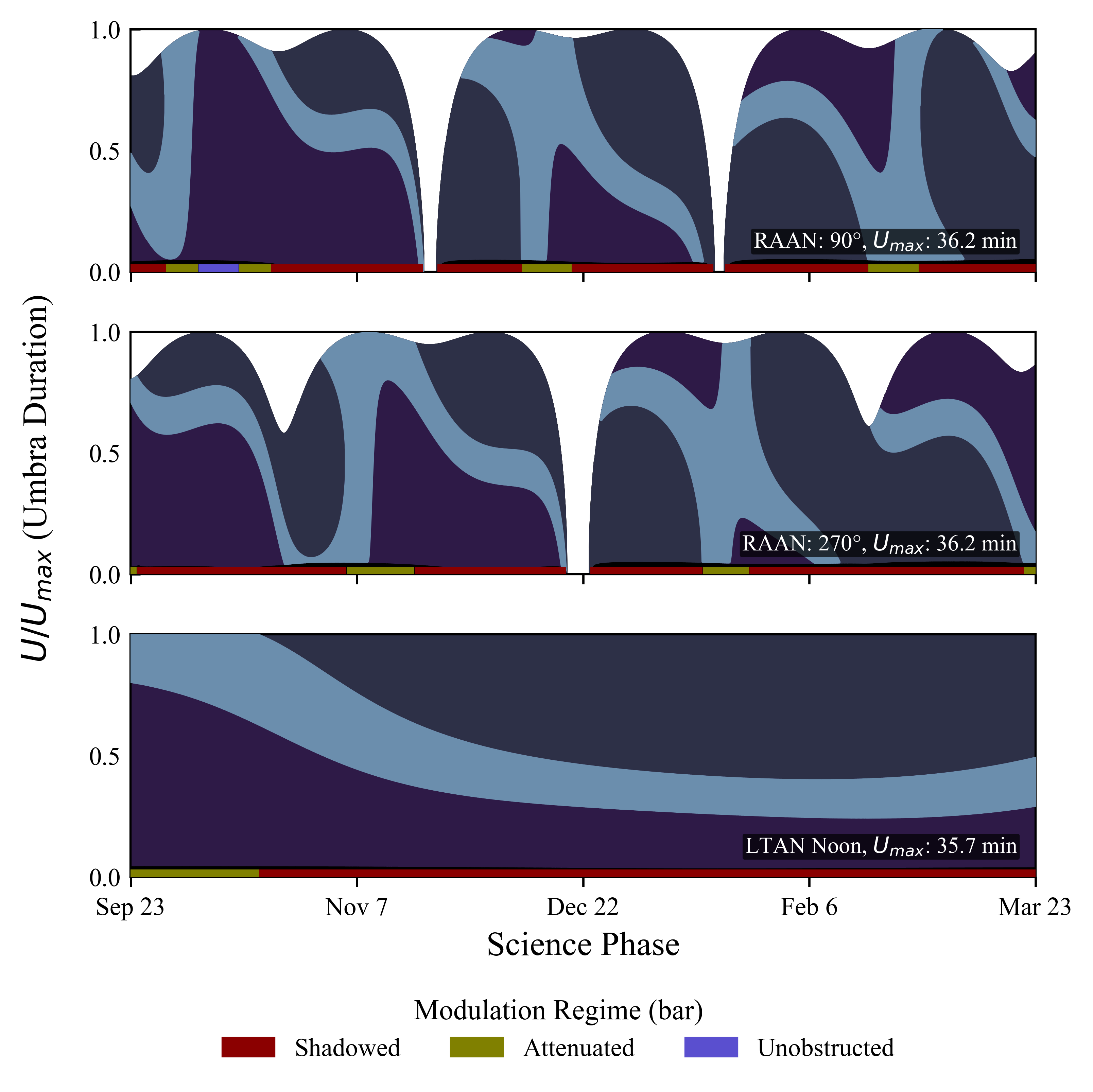}
\caption{\textbf{Cygnus:} Simulated obstruction landscape for strongly interacting DM flux targeting and modulation studies during the 2026 Autumnal–2027 Vernal Equinoxes, across mid-inclination and Noon LTAN SSO domains. %Continuous shading follows the convention shown in Fig.~\ref{fig:GCdomains}.
Observation regimes are classified by the Boolean bar (Crimson: shadowed, Olive: attenuated, Iris: unobstructed). Persistent shadowing throughout winter supports high-frequency sampling of Earth-modulated DM flux.}
\label{fig:CYGNUSdomains}
\end{figure}

In mid-inclination deployments, unobstructed windows diminish from mid-spring through winter as the umbra axis diverges from the Cygnus vector. This geometry produces extended shadowed regimes, where Earth physically obstructs the incoming DM flux over a passage segment. Attenuated conditions arise when the Earth's limb and the target remain within the FOV, receiving a reduced flux. As shown in Figures~\ref{fig:VACYGNUSdomains} and~\ref{fig:CYGNUSdomains}, shadowed regimes often contain a sequence of unobstructed, partially, and completely obstructed segments within a single passage. This segmentation enables observations within shadowed windows to receive exposure to a fully modulated DM flux profile.

%------------------- Shifting Landscape Figures -----------------

The SSO domain supports Cygnus observations through a frequently shadowed landscape, as reported in Table~\ref{tab:CYGNUSsummary}. This environment provides a well-structured modulating mechanism throughout the year, with autumn months offering attenuated-only windows (free of shadowing). Although unobstructed access is rare (passage without obstruction), the high frequency of shadowed and attenuated windows support dense temporal sampling across the distinct obstruction regimes, potentially resolving seasonal sub-dominant trends in the modulated DM flux~\citep{emken:JCAP2019}.

The obstruction analysis maps these regimes into scheduling opportunities. For Sagittarius, operations concentrate between late spring and mid-summer, when observation best aligns toward the Galactic Center (favoring unobstructed windows). For Cygnus, observations exploit the occurrence of attenuated and shadowed segments within passages. When Sagittarius targeting becomes inaccessible due to the closure of useful observation windows, objectives targeting Cygnus can be prioritized.

This dual-objective framework supports continuous science operations by selecting observation windows according to the landscape. The operations plan will be constructed from Day-of-Flight orbit parameters by re-simulating the obstruction landscape and incorporating telemetry updates for daily scheduling during DarkNESS science operations.

\section{Thermal Considerations}
\label{sec:thermal}

The MCM (Fig.~\ref{fig:instrument}) operates within a bounded thermal range of $170 \pm 5\,\mathrm{K}$ to suppress dark current~\citep{Widenhorn_2002}. The detector’s thermal inertia (60\,g) results in a slow thermal response, with several hours required to chill the detector module. Consequently, the MCM must remain thermally conditioned throughout each orbit cycle. This continuous requirement constrains the thermal architecture. Passive elements such as phase change materials are not well-suited for continuous-load applications, as they require recharge cycling to regenerate latent heat capacity~\citep{Elshaer2023PCM}. Active regulation is therefore implemented using the cryocooler~\citep{Mok2020_K508N}, complemented by multilayer insulation (MLI) to reduce radiative loading on the MCM. 

\begin{figure}[t]
\centering
\includegraphics[width=0.8\columnwidth]{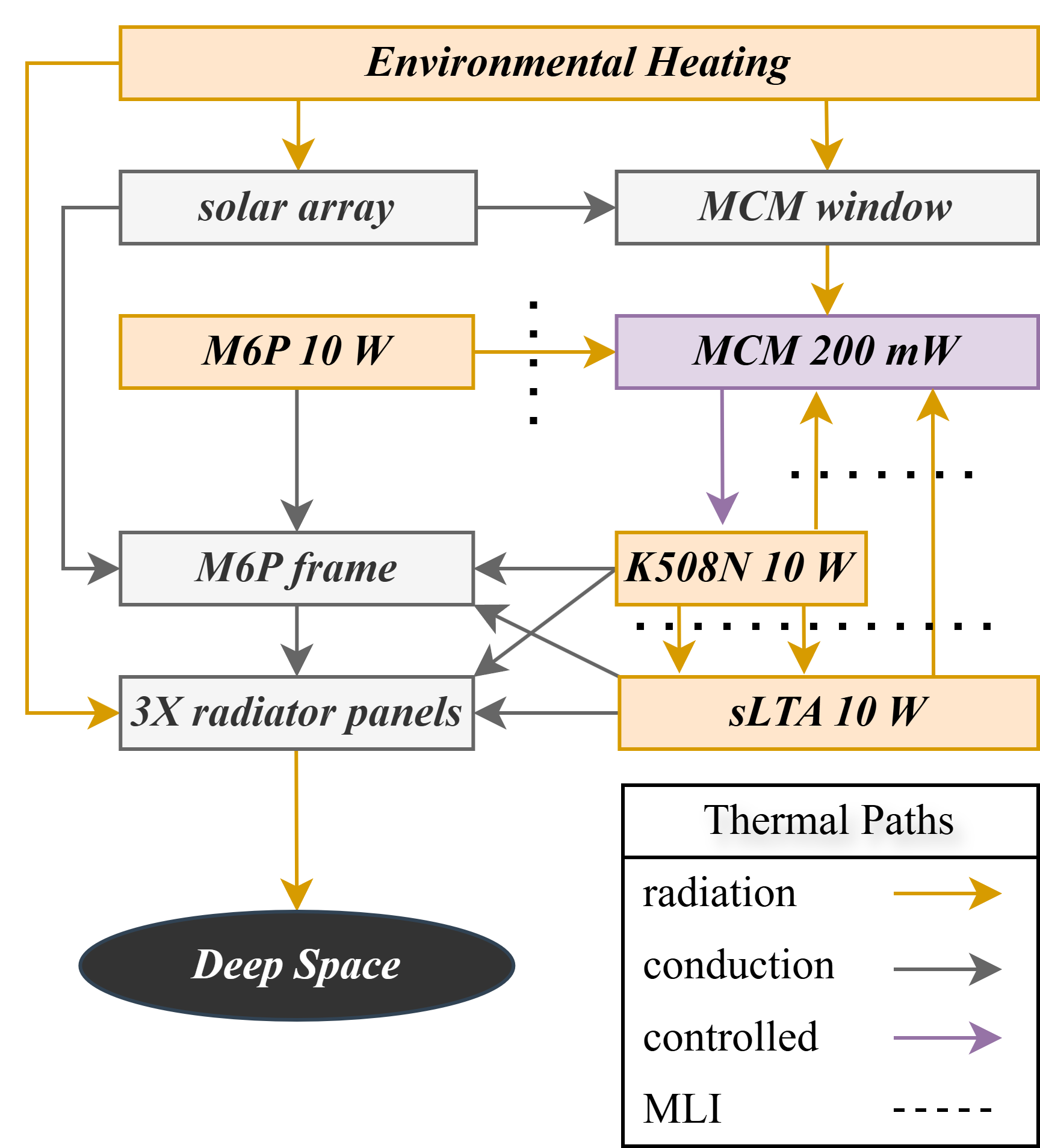}
\caption{Thermal control architecture of the DarkNESS instrument. The Ricor K508N integral rotary cryocooler regulates the MCM cold-end via a conductive copper bracket, while passive copper thermal straps route waste heat from the cryocooler and sLTA electronics to body-mounted radiators. Multi-layer insulation (MLI) reduces parasitic heat load from internal electronics and external environmental heating, mitigating regenerator conditions critical to managing the MCM at $170\pm5$\,K.}
\label{fig:thermal_architecture}
\end{figure}

Passive thermal elements are necessary to support the continuous operation of the cryocooler’s reversed Stirling cycle, where the cooling power depends on maintaining a sensitive temperature differential across the regenerator~\citep{huang2023stirling}. The thermodynamic efficiency of this cycle, and the regenerator in particular, is sensitive to parasitic heat loads arriving through the MCM interface and cryocooler body~\citep{Grossman2012_TransientCryocooler}. Therefore, custom passive elements are interfaced to key cryocooler dissipating surfaces to offload waste heat that influences regenerator behavior under transient conditions. This integrated thermal configuration overview is shown in Fig.~\ref{fig:thermal_architecture}.

A copper bracket couples the cryocooler’s cold fingertip to the MCM substrate, while auxiliary copper thermal straps route waste heat from the compressor, expander, and motor housing to body-mounted radiator panels. Coated with Socomore Aeroglaze A276 white thermal paint, these panels radiatively exchange waste heat toward deep space while suppressing spectral absorption. By managing the surface temperatures surrounding the regenerator, these passive paths preserve the cyclic temperature gradient essential for adequate conditioning of the MCM during orbit-wise observation cycles and removing fluctuating parasitic heat loads.

Further thermal stability is achieved by placing multi-layer insulation (MLI), which mitigates the parasitic heat loads by shielding radiative exposure from internal warm payload and subsystem components. In particular, the sLTA electronics generate continuous waste heat during observation cycles (consuming 10\,W), contributing a dominant internal heat source in close proximity to the MCM. 

The MLI and radiator-cooled thermal paths collectively stabilize the regenerator boundary conditions, helping to maintain the cold-end temperature gradient required for the reversed Stirling cycle. This integrated thermal control architecture manages the transient balance between cryocooler performance, internal dissipation, and environmental heating, enabling low-noise operation of the MCM within the constraints of the 6U platform.

\subsection{Active Thermal System}

\begin{figure}[t]
    \centering
    \includegraphics[width=1\columnwidth]{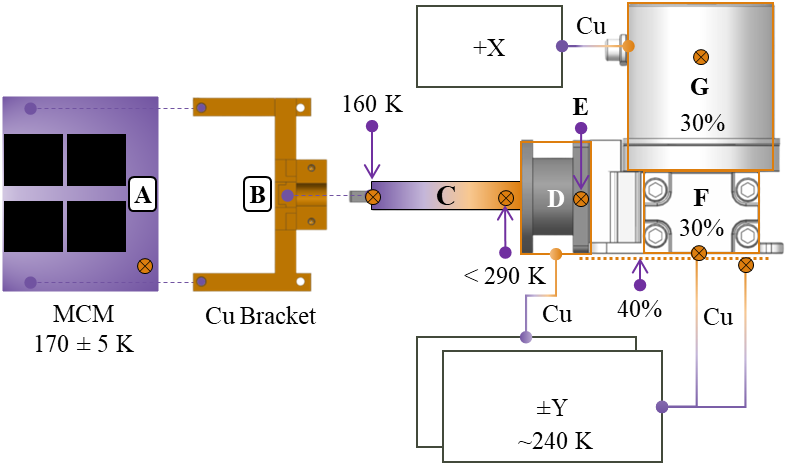}
    \caption{\textbf{Active Thermal Control:} The Ricor K508N cryocooler actively maintains the MCM substrate (A) with a 100\% duty-cycle and is interfaced by a copper thermal bracket (B). The fixed-point closed loop control regulates the motor current to maintain a resister-set temperature at the cold fingertip (C). The thermal gradient along the cold finger is best managed by offloading the flange (D) due to its proximity to the expander region (E). The thermal condition of these surfaces impedes the performance of the regenerator cycle. Care must be taken to off-load the compressor cover (F), the motor assembly (G), and the bottom mounting surface. The thermal dissipation through these surfaces is a percentage of the total power consumption. Notable regions for temperature monitoring are marked with targets (Orange), and desired operating temperatures as configured for DarkNESS are labeled.}
    \label{fig:cryocooler_MCM}
\end{figure}

The cryocooler operates continuously in a closed-loop, fixed-setpoint control configuration to maintain the MCM at $170\,\mathrm{K}$. The cryocooler’s mechanical coupling and thermal control interfaces are shown in Fig.~\ref{fig:cryocooler_MCM}, where key components are labeled for clarity. A Lakeshore DT-670-BO temperature diode monitors the cold fingertip (C), while an in-line variable resistor sets the control point by regulating the bias voltage delivered to the motor assembly controller (G). The resistor value required for a $170\,\mathrm{K}$ setpoint under laboratory conditions is $326\,\Omega$, based on the diode’s voltage–temperature calibration. In-flight tuning allows adjustment of this setpoint within an $800$--$1000\,\Omega$ range to compensate for transient parasitic heat loads received by the MCM. This configuration mitigates orbit-wise thermal variability and defines a bounded operational condition monitored throughout the mission. Maintaining the thermal setpoint is a key aspect of the Observatory's operation strategy, because temperature deviations at the cold fingertip affect the MCM's susceptibility to dark current fluctuations.

The cryocooler operates in commanded voltage modes ($12\,\mathrm{V}$ default; $17\,\mathrm{V}$ and $20\,\mathrm{V}$ selectable) that determine both the rate and magnitude of heat removal. Higher voltages allow the system to respond to larger or more rapidly varying thermal loads, such as those induced by indirect spectral heating during sunlit passage. This increase in cooling capacity also raises electrical power draw and waste heat dissipation at key interfaces, which are managed through passive thermal elements (labeled in Fig~\ref{fig:cryocooler_MCM}). The effectiveness of this offloading depends on the Observatory’s orientation, prompting the use of secondary pointing mode constraints described in Sec.~\ref{sec:configuration} to maintain favorable radiative efficiency to deep space. These constraints reduce reliance on sustained high-voltage operation, which is challenging to maintain for prolonged durations. The cryocooler’s cooling capacity spans $1000$ to $1300\,\mathrm{mW}$, with power draw peaking near $10\,\mathrm{W}$ during chill-down at $17\,\mathrm{V}$ and leveling below $5\,\mathrm{W}$ during fixed-point operation at $12\,\mathrm{V}$.

The cryocooler interfaces directly to the MCM via a copper bracket mounted to the cold fingertip, shown in Fig~\ref{fig:cryocooler_MCM}. Laboratory tests indicate a chill-down time of approximately three hours for the $\sim$60~g ceramic and silicon MCM assembly. During operation, the four wire-bonded skipper-CCDs contribute a total thermal load of 200~mW onto the substrate. The large surface area of the aluminum nitride substrate ($\sim$38~cm\textsuperscript{2}) increases radiative coupling, requiring stable boundary conditions for the cold interface.

\subsection{Passive Thermal System}
\begin{figure}[t]
    \centering
    \includegraphics[width=0.85\columnwidth]{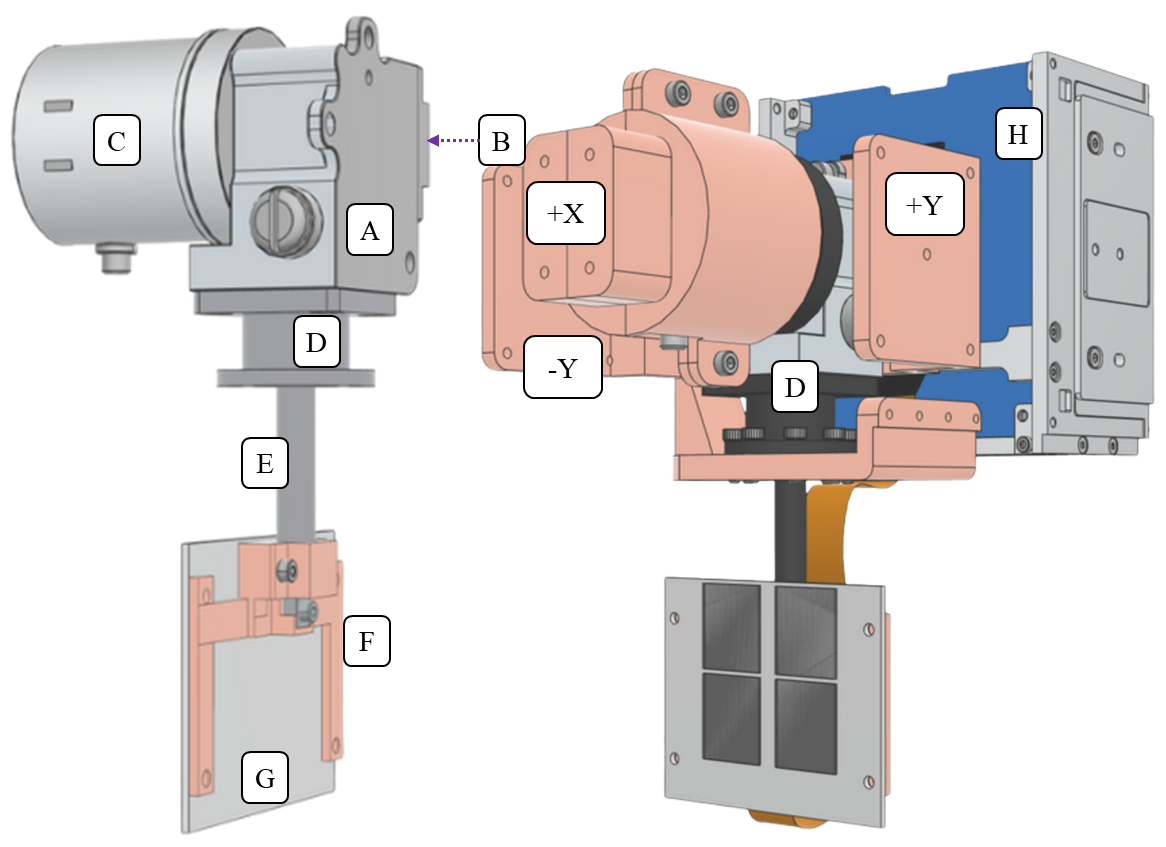}
    \caption{\textbf{Passive Thermal Control:}
    CAD model of the cryocooler and MCM assembly. \textbf{Left:} Key heat-dissipating surfaces include the mounting base (A), compressor cover (B), and motor housing (C). The flange (D) can be used to offload the expander, improving the gradient along the cold finger (E), which interfaces with the MCM (G) via a copper bracket (F). \textbf{Right:} The cryocooler is mounted to a custom Copper bracket at the mounting surface (A) and flange (D). A flex cable links the MCM (G) to the sLTA (H). Cryocooler waste heat from (A,B,D) is transported via the Copper bracket to the ($\pm Y$) radiator panels, while the motor cover (C) is offloaded to the ($+X$) radiator.}
    \label{fig:payload_cad}
\end{figure}

Passive thermal management addresses waste heat from three key cryocooler surfaces labeled in Fig.~\ref{fig:payload_cad}: the mounting base (A), compressor cover (B), and motor housing (C). These contribute approximately 40\%, 30\%, and 30\% of the input power dissipation, respectively. Thermal transport from these surfaces is handled via copper brackets that clamp to the body-mounted, painted radiator panels ($\pm Y$). The motor housing is preferentially offloaded to the ($+X$) radiator panel, maintaining continuous exposure towards deep space during sunlit passage (defined by the pointing modes, see Fig.~\ref{fig:sunlit} for concept cartoon).

The sLTA (label H in Fig.~\ref{fig:payload_cad}) readout electronics generate $10\,\mathrm{W}$ when powered for observations during umbral passage. Passive thermal management is accomplished by a custom thermal board box developed at UIUC. It features a sliding-wall feature that clamps to the thermal edge ring around the perimeter of each circuit board. The alodine-coated aluminum walls interface directly to the $\pm$Y radiator panels and manage the hottest component (1000\,MB Ethernet chip) below $55^\circ$\,C.

\subsection{Laboratory Testing}

\begin{figure}[t]
    \includegraphics[width=0.9\columnwidth]{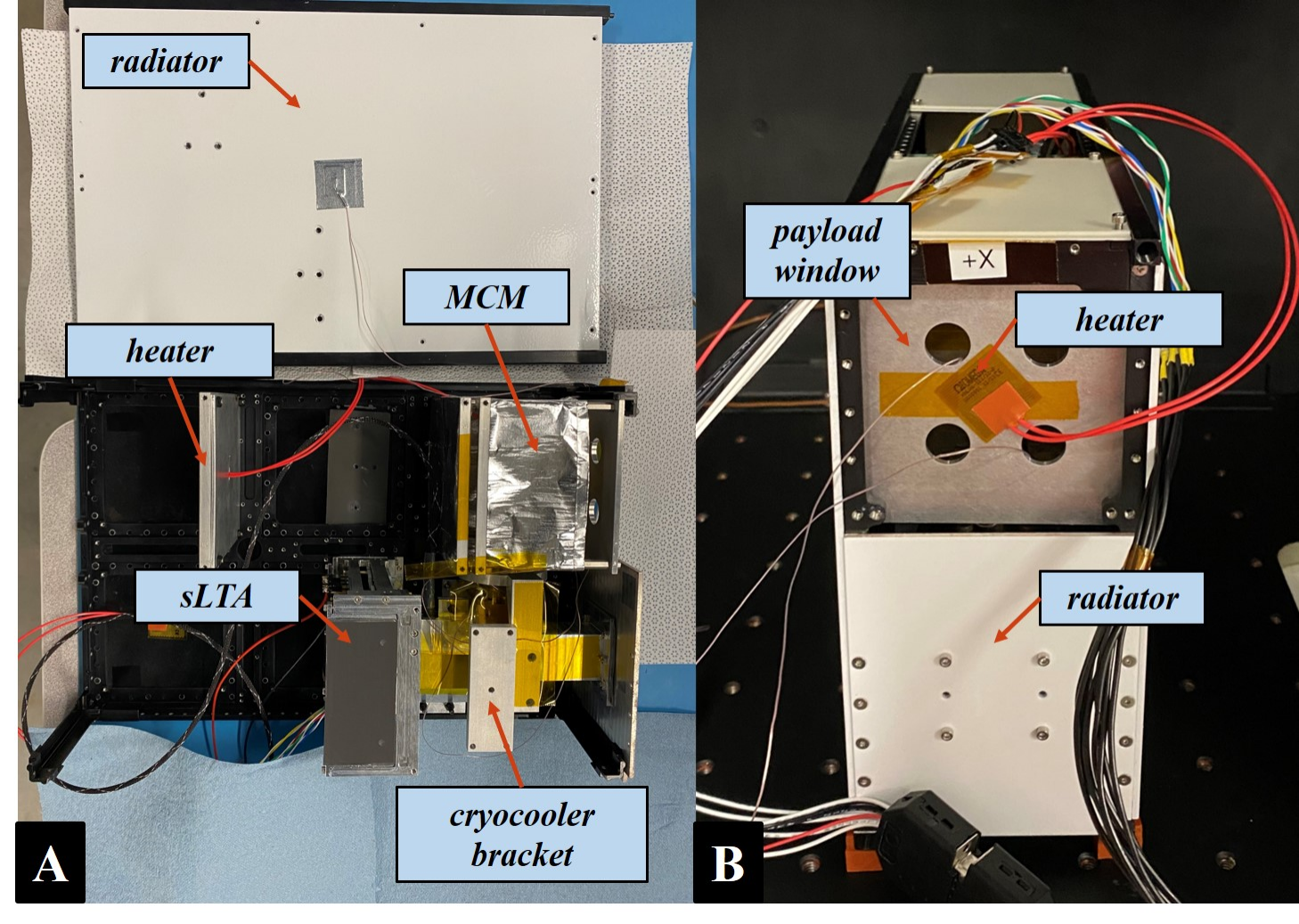}
    \centering
    \caption{\textbf{A} Design of the DarkNESS thermal system configuration. The 6U NanoSatellite has dimensions 30\,cm x 20\,cm x 10\,cm. \textbf{B} Engineering model used at UIUC to test the thermal design for DarkNESS. The model has the cryocooler, readout electronics (sLTA), detector (MCM) assembly (with dummy sensors), and radiator panels with surface treated using Socomore Aeroglaze A276 white paint.\label{fig:thermaltest}}   
\end{figure}

Hardware testing of the thermal control system was performed as part of early risk-mitigation efforts. A flight-worthy NanoSatellite frame was acquired from KNA for prototype payload integration and testing under thermal vacuum (TVAC) conditions. A mechanical MCM featuring four skipper-CCDs mounted to the ceramic substrate was used alongside a functional sLTA housed in an aluminum board box for thermal management. 
The cryocooler was mounted using a placeholder thermal bracket design, with its cold fingertip interfacing the MCM substrate via a copper bracket and an indium foil thermal interface. The payload and thermal hardware components are pictured in Fig.~\ref{fig:thermaltest}.

Thermocouples were attached to key surfaces (described by the orange targets illustrated in Fig.~\ref{fig:cryocooler_MCM}) of the cryocooler, MCM ceramic, and radiator panels to observe the thermal response. Two Kapton heaters were also integrated into the experiment to simulate external heating of the MCM window (set at 2\,W) and the internal NanoSatellite subsystem components (set at 7\,W). Thermal testing was performed inside the TVAC equipped with a liquid nitrogen shroud, which was brought to a high vacuum using a roughing and turbo pump setup.

\begin{figure}[h]
    \includegraphics[width=\columnwidth]{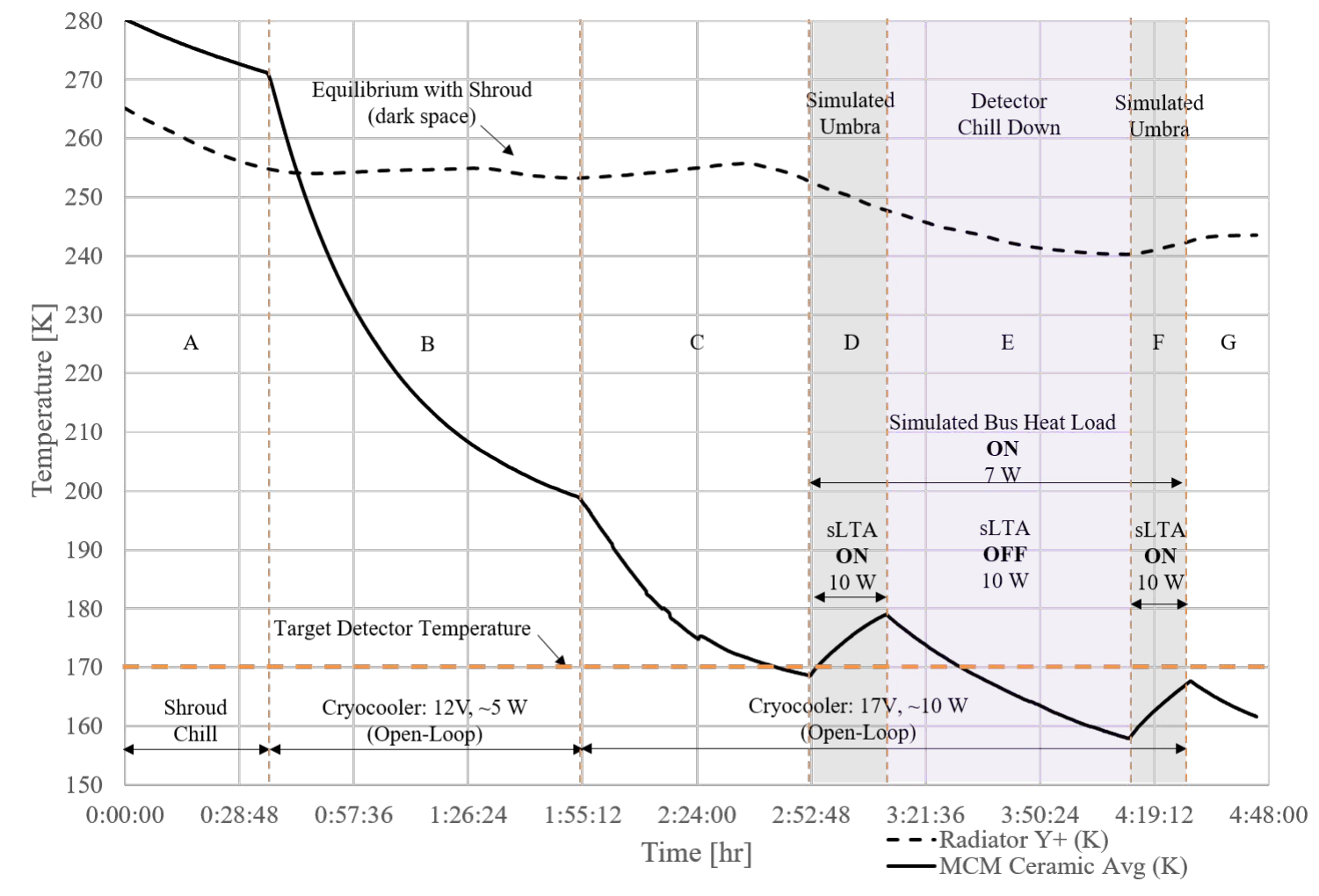}
    \centering
    \caption{Results from TVAC tests at LASSI-UIUC. The chamber was maintained at $5.0\times 10^{-4}$\,Torr and the shroud at 255\,K (Stage A). The cryocooler (10 W) gets the MCM to 168\,K (Stage C). The bus heat load is added (7\,W), and the cycles of 15-minute sLTA readout with the additional load (10\,W) are shown in Stages D and F. The MCM stays around the target operating point 170\,K. These tests were done with open-loop control for the cryocooler. Closed-loop cryocooler control will be implemented to reduce the temperature fluctuations around the operating point. The cryocooler is designed to operate for 20,000 hours, and its operational lifetime is not a limiting factor for the mission~\citep{Regev_2016}.\label{fig:thermalResults2024}}
\end{figure}

The thermal response of the MCM was evaluated in a five-hour thermal vacuum test, with results shown in Fig.~\ref{fig:thermalResults2024}. The average temperature of the aluminum nitride substrate was monitored across sequential test phases simulating orbital thermal conditions and power cycling.

The chamber was cooled to approximately 255\,K using the liquid nitrogen shroud prior to cryocooler activation. In Phase B, the cryocooler was powered at 12\,V, initiating cool-down from ambient. The chamber temperature was manually regulated and remained warmer than the expected in-orbit environment. In Phase C, the cryocooler voltage was increased to 17\,V, allowing the MCM to reach the target 170\,K setpoint.

In Phase D, the sLTA electronics were activated with a 10\,W load, introducing internal radiative heating and causing a measurable temperature rise at the MCM substrate. During Phase E, the sLTA was powered down, and a 7\,W avionics heater was applied to simulate bus-generated heat. The cryocooler reduced the MCM temperature below 170\,K under these conditions. In Phase F, the sLTA was reactivated and a 1\,W heater simulating detector window loading was applied. The MCM remained below 170\,K for at least 15 minutes, replicating a full-duration umbral science observation cycle. The chamber shroud temperature further decreased to 240\,K to simulate deeper umbral-like conditions but remained above true orbital minima. The test concluded in Phase G with the depletion of the liquid nitrogen supply.

This preliminary TVAC test demonstrates that the DarkNESS thermal architecture can sustain the MCM at 170\,K for 15-minute intervals under simulated orbital conditions. Based on the results, the design was modified to reduce the MCM’s thermal mass and isolate it from the radiative loading of nearby electronics. A complete TVAC test campaign is ongoing at FNAL to evaluate the flight thermal control system using the integrated MCM and cryocooler operating in a fixed-setpoint control mode.
This campaign includes representative internal and external heat loads expected during science operations in LEO. Future work will present the full results from these laboratory tests.

\section{Conclusions and Future Work}
\label{sec:conclusion}

The DarkNESS mission integrates a skipper-CCD instrument with a compact sLTA readout electronics and a closed-loop cryocooler in a thermally-conditioned 6U NanoSatellite to search for DM in LEO. The mission targets two DM detection signatures--keV X-rays from decaying DM and low-energy ionization events from strongly interacting DM. The Observatory will conduct umbral observations, with the imaging cadence shaped by celestial dynamics and bounded background considerations. The system design reflects the constraints imposed by the science objectives and payload requirements, with orbit analysis and thermal testing confirming operational feasibility under those conditions and identifying observation windows sufficient to meet mission goals.

Current efforts focus on thermal system validation and integrated testing. A full thermal vacuum campaign at Fermilab evaluates flight-version hardware—including the MCM, cryocooler, and sLTA electronics—under representative environmental loading. The testbed couples the payload to non-flight bus avionics through a KNA-built FlatSat interface, enabling payload software development and hardware-in-the-loop testing. In-chamber characterization uses a \textsuperscript{55}Fe source to assess skipper-CCD performance in simulated space conditions. Concurrent software development establishes end-to-end control of the payload from the KNA interface and defines the operational baseline for on-orbit commanding and data acquisition. Design, testing, and performance of the flight-configuration thermal control system for DarkNESS will be reported upon completion of the test campaign.

\section*{Acknowledgments}
\label{sec:acknowledgement}

We are grateful for the support of the Heising-Simons Foundation under Grant No. 2023-4612. This work is supported by the Fermilab Laboratory Directed Research and Development (LDRD) program. Fermilab operates under U.S. Department of Energy (DOE) contract No. DE-AC02-07CH11359. PA acknowledges support from the DOE Office of Science Graduate Student Research (SCGSR) Program under solicitation DE-FOA-0002999 in the High Energy Physics research area, which supported doctoral research contributions at Fermilab and the University of Illinois Urbana-Champaign. The authors also acknowledge Firefly Aerospace for supporting the University of Illinois student team through the award of a DREAM~2.0 launch opportunity.
RE acknowledges support from DOE Grant DE-SC0025309, Simons Investigator in Physics Award~623940, Heising-Simons Foundation Grant No.~79921, Binational Science Foundation Grant No.~2020220.  
HX is supported in part by DOE Grant DE-SC0009854, Simons Investigator in Physics Award~623940, and the Binational Science Foundation Grant No.~2020220.

\addtocontents{toc}{\protect\setcounter{tocdepth}{1}}

\section*{}\label{sec:references}\addcontentsline{toc}{section}{References}

\onecolumn{
\bibliographystyle{JHEP}
\bibliography{main,main-phi}

}

\clearpage

\end{document}